\documentclass[11pt]{article}
\usepackage{amsfonts,amsmath,amssymb}
\usepackage{enumerate}
\usepackage{hyperref}
\usepackage{bbm}
\usepackage{nicefrac}
\usepackage[all]{xy}
\usepackage{graphicx}
\usepackage{bm}
\usepackage{makecell}
\usepackage[table]{xcolor}
\usepackage{scalerel}

\usepackage{upgreek}

\usepackage{booktabs}

\newcommand{\be}{\begin{equation}}
\newcommand{\ee}{\end{equation}}

\newcommand{\bea}{\begin{eqnarray}}
\newcommand{\eea}{\end{eqnarray}}

\newcommand{\bes}{\begin{subequations}}
\newcommand{\ees}{\end{subequations}}

\newcommand{\cN}{{\cal N}}
\newcommand{\cA}{{\cal A}}

\newcommand{\tr}{\mbox{tr}}

\def\sst#1{{\scriptscriptstyle #1}}

\def\0{{\sst{(0)}}}
\def\1{{\sst{(1)}}}
\def\2{{\sst{(2)}}}
\def\3{{\sst{(3)}}}
\def\4{{\sst{(4)}}}
\def\5{{\sst{(5)}}}
\def\6{{\sst{(6)}}}
\def\7{{\sst{(7)}}}
\def\8{{\sst{(8)}}}

\def\cA{{{\cal A}}}

\usepackage{multirow}
\usepackage{rotating}

\newcommand{\ba}{\begin{align}}
\newcommand{\ea}{\end{align}}

\newcommand{\bse}{\begin{subequations}}
\newcommand{\ese}{\end{subequations}}

\allowdisplaybreaks

\begin{document}

\makeatletter
\renewcommand{\theequation}{\thesection.\arabic{equation}}
\@addtoreset{equation}{section}
\makeatother

\begin{titlepage}

\begin{flushright}
IFT-UAM/CSIC-19-108 \\
HIP-2019-24/TH
\end{flushright}

   \begin{center}
   \baselineskip=16pt

\begin{Large}\textbf{
Halving ISO(7) supergravity
}\end{Large}

\vspace{40pt}

{\large  \textbf{Adolfo Guarino}$^{a, b}$ ,\,\, \textbf{Javier Tarr\'io}$^{c}$ \,\,and\,\,  \textbf{Oscar Varela}$^{d, e}$}

\vspace{25pt}

$^a$\,{\normalsize  
Departamento de F\'isica, Universidad de Oviedo,\\
Avda. Federico Garc\'ia Lorca 18, 33007 Oviedo, Spain.}
\\[5mm]

$^b$\,{\normalsize  
Instituto Universitario de Ciencias y Tecnolog\'ias Espaciales de Asturias (ICTEA) \\
Calle de la Independencia 13, 33004 Oviedo, Spain.}
\\[5mm]

$^c$\,{\normalsize  Department of Physics and Helsinki Institute of Physics \\ P.O.Box 64
FIN-00014, University of Helsinki, Finland.}
\\[5mm]

$^d$\,{\normalsize  
Department of Physics, Utah State University, Logan, UT 84322, USA.}
\\[5mm]

$^e$\,{\normalsize  
Departamento de F\'\i sica Te\'orica and Instituto de F\'\i sica Te\'orica UAM/CSIC, \\  Universidad Aut\'onoma de Madrid, Cantoblanco, 28049 Madrid, Spain.}
\\[10mm]

\mbox{\texttt{adolfo.guarino@uniovi.es} \, , \, \texttt{javier.tarrio@helsinki.fi} \, , \, \texttt{oscar.varela@usu.edu}}

\vspace{20pt}


\end{center}

\begin{center}
\textbf{Abstract}
\end{center}

\begin{quote}

Half-maximal, $\,\mathcal{N}=4\,$, sectors of $\,D=4\,$ $\,\mathcal{N}=8\,$ supergravity with a dyonic ISO(7) gauging are investigated. We focus on a half-maximal sector including three vector multiplets, that arises as a certain $\,\textrm{SO}(3)_{\textrm{R}}$-invariant sector of the full theory. We discuss the embedding of this sector into the largest half-maximal sector of the $\,\mathcal{N}=8\,$ supergravity retaining six vector multiplets. We also provide its canonical $\,\mathcal{N}=4\,$ formulation and show that, from this perspective, our model leads in its own right to a new explicit gauging of $\,\mathcal{N}=4\,$ supergravity. Finally, expressions for the restricted duality hierarchy are given and the vacuum structure is investigated. Five new non-supersymmetric AdS vacua are found numerically. The previously known $\,\mathcal{N}=2\,$ and $\,\mathcal{N}=3\,$ AdS vacua are also contained in our $\,\mathcal{N}=4\,$ model. Unlike when embedded in previously considered sectors with fewer fields, these vacua exhibit their full $\,\mathcal{N}=2\,$ and $\,\mathcal{N}=3\,$ supersymmetry within our $\,\mathcal{N}=4\,$ model.

\end{quote}

\vfill

\end{titlepage}

\tableofcontents

\hrulefill
\vspace{10pt}

\section{Motivation}

Gauged supergravities in lower dimensions that descend consistently from string or M-theory, and whose scalar potentials possess anti-de Sitter (AdS) vacua, prove to be extremely helpful venues to study aspects of large-$N$ realisations of the AdS/CFT correspondence \cite{Maldacena:1997re}. Out of these, the handful of gauged supergravities with a {\it maximal} amount of supersymmetry and specific gauge groups enjoy an even more distinguished status, as they are related to the AdS/CFT correspondences associated with the standard half-BPS conformal branes. This is the case for the maximal supergravities in $D=7$, $D=5$ and $D=4$ with gauge groups SO(5) \cite{Pernici:1984xx}, SO(6) \cite{Gunaydin:1984qu} and SO(8) \cite{deWit:1982ig}: they are respectively related holographically to the superconformal field theories, and mass deformations thereof, defined on the M5, the D3 \cite{Maldacena:1997re} and the M2 \cite{Aharony:2008ug} branes. 

Recently, $D=4$ $\cN=8$ supergravity with gauge group $\textrm{ISO}(7) = \textrm{SO}(7) \ltimes \mathbb{R}^7$, has also been shown to be a member of the select club of maximal supergravities with a holographic interpretation \cite{Guarino:2015jca}, at least when ISO(7) is gauged dyonically in the sense of \cite{Dall'Agata:2012bb,Dall'Agata:2014ita,Inverso:2015viq}. The purely electric ISO(7) supergravity \cite{Hull:1984yy} descends consistently \cite{Hull:1988jw} from type IIA \cite{Giani:1984wc}, but it lacks AdS (or dS) vacua \cite{DallAgata:2011aa}. Similarly, the dyonically-gauged version \cite{Dall'Agata:2012bb,Dall'Agata:2014ita,Inverso:2015viq} of ISO(7) supergravity \cite{Guarino:2015qaa} descends consistently \cite{Guarino:2015jca,Guarino:2015vca} (see also \cite{Ciceri:2016dmd,Cassani:2016ncu,Inverso:2016eet}) from mass-deformed type IIA supergravity \cite{Romans:1985tz} but, in contrast to its purely electric counterpart \cite{Hull:1984yy}, its scalar potential does attain AdS extrema \cite{DallAgata:2011aa,Borghese:2012qm,Gallerati:2014xra,Guarino:2015qaa}. These vacua preserve different fractions $\cN < 8$ of supersymmetry and uplift to fully-fledged AdS$_4$ solutions of massive type IIA supergravity \cite{Guarino:2015jca,Varela:2015uca,Pang:2015vna,DeLuca:2018buk}. The latter are dual \cite{Guarino:2015jca} to three-dimensional superconformal field theories of the simple class discussed in \cite{Schwarz:2004yj,Gaiotto:2007qi}, and arise as infrared Chern-Simons phases of the D2-brane field theory \cite{Guarino:2016ynd}. The field theory spectra have been partially determined holographically \cite{Pang:2015rwd,Pang:2017omp}, and solutions of various types have been constructed in the gauged supergravity. These include domain-walls \cite{Guarino:2016ynd,Bobev:2018ugk}, defects \cite{Suh:2018nmp} or black holes, both the full asymptotically-AdS geometries \cite{Guarino:2017eag,Hosseini:2017fjo} and their near-horizon regions \cite{Guarino:2017pkw}. These solutions are all in perfect agreement with the corresponding field theory predictions. For example, the entropy of those black holes has been succesfully matched from the field theory \cite{Azzurli:2017kxo,Hosseini:2017fjo,Benini:2017oxt,Liu:2018bac}. Further aspects of these AdS$_4$/CFT$_3$ dualities have been explored in \cite{Fluder:2015eoa,Araujo:2016jlx,Araujo:2017hvi}.

Other than the fact that dyonic ISO(7) supergravity pertains to the D2-brane and the latter does not support a maximally supersymmetric conformal field theory on its worldvolume, everything else works as for the maximal supergravities \cite{Pernici:1984xx,Gunaydin:1984qu,deWit:1982ig} relevant for the usual AdS/CFT descriptions of the superconformal, M5, D3 and M2, branes. In fact, as for the latter, better-known superconformal cases, there are aspects that make these new AdS$_4$/CFT$_3$ dualities rather unique, both from the supergravity and from the field theory sides. On the one hand, it is now established that $D=4$ $\cN=8$ gauged supergravities tend to come in families whose members have the same gauge group but different electric/magnetic duality frames prior to introducing the gauging \cite{Dall'Agata:2012bb,Dall'Agata:2014ita,Inverso:2015viq}. For ISO(7) supergravity, this electric/magnetic deformation is still compatible with a higher-dimensional origin, the feature which ultimately renders it relevant for top-down holography. In contrast, the analogue deformation of the SO(8) gauging \cite{Dall'Agata:2012bb} cannot be consistently uplifted \cite{deWit:2013ija,Lee:2015xga} to conventional $D=11$ supergravity \cite{Cremmer:1978km} unlike, as is well-known \cite{deWit:1986iy}, the purely electric representative \cite{deWit:1982ig} in that class. On the other hand, one may envisage large classes of three-dimensional superconformal field theories in the simple class considered in \cite{Schwarz:2004yj,Gaiotto:2007qi} that differ in their matter couplings and flavour symmetries. However, weakly coupled AdS$_4$ gravity duals are ruled out for most of those, as their spectra tend to exhibit light operators with unbounded spin and expotential growth \cite{Minwalla:2011ma}. The handful of cases not ruled out by \cite{Minwalla:2011ma} turn out to be the ones relevant for the AdS$_4$/CFT$_3$ dualities of \cite{Guarino:2015jca}. 

For these reasons, it is interesting to investigate further aspects of $D=4$ $\cN=8$ dyonic ISO(7) supergravity from an intrinsically four-dimensional perspective. A useful way to do this is to truncate the theory to a more manageable sector with fewer fields, where a concrete parameterisation for the scalar manifold can be introduced. This allows one to obtain explicit expressions for the scalar potential and therefore study its vacuum structure. It also allows one to obtain concrete expressions for the tensor \cite{deWit:2008ta,deWit:2008gc} and duality hierarchies \cite{Bergshoeff:2009ph}, which eventually play a significant role in the type IIA uplift \cite{Guarino:2015vca}. In this spirit, further $\cN=1$ or $\cN=2$ subsectors of the full $\cN=8$ ISO(7) theory have already been presented in \cite{Guarino:2015qaa,Kim:2018sdw}. 

In this work, we focus on half-maximal, $\cN=4$, sectors of ISO(7) supergravity. We find that $\cN=4$ sectors provide an excellent compromise between  particularity and generality. On the one hand, $\cN=4$ sectors are small enough to admit a tractable parameterisation in terms of explicit scalar fields, unlike the full $\cN=8$ theory. On the other hand, $\cN=4$ sectors are large enough to contain relevant features of ISO(7) supergravity in a unified way, which is still simplified with respect to their description within the full $\cN=8$ theory. For example, the SU(3) and  $\textrm{SO}(3)_{\textrm{d}} \times \textrm{SO}(3)_{\textrm{R}}$ invariant subsectors constructed in \cite{Guarino:2015qaa} respectively contain, and mutually exclude, the $\cN=2$ \cite{Guarino:2015jca} and $\cN=3$ \cite{Gallerati:2014xra} AdS vacua of ISO(7) supergravity. Moreover, the latter manifests itself as $\cN=0$ within the $\textrm{SO}(3)_{\textrm{d}} \times \textrm{SO}(3)_{\textrm{R}}$ sector and only when embedded in the full $\cN=8$ theory becomes its actual $\cN=3$ supersymmetry apparent. The reason for this awkward, yet perfectly reasonable, feature is that the $\cN=3$ gravitini that remain massless at that vacuum transform as a triplet of the residual R-symmetry group, $\textrm{SO}(3)_{\textrm{d}}$, and are thus truncated out from the $\textrm{SO}(3)_{\textrm{d}} \times \textrm{SO}(3)_{\textrm{R}}$-singlet sector. 

More concretely, we explicitly construct, in section~\ref{sec:embedding_N8}, the explicit Lagrangian of a subsector of $\cN=8$ ISO(7) supergravity that corresponds to $\cN=4$ supergravity coupled to three vector multiplets. This sector is invariant under a certain $\textrm{SU}(2)  \sim \textrm{SO}(3)$ subgroup of $\textrm{SO}(7) \subset \textrm{ISO}(7)$. This SU(2) is the subgroup of the SU(3) residual symmetry group of the $\cN=2$ point such that $\bm{3} \rightarrow \bm{2} +\bm{1}$. It also coincides with the $\textrm{SO}(3)_{\textrm{R}}$ factor of the $\textrm{SO}(3)_{\textrm{d}} \times \textrm{SO}(3)_{\textrm{R}}$ invariance group of the $\cN=3$ point. The $\textrm{SO}(3)_{\textrm{R}}$-invariant sector thus contains, by construction, both $\cN=2$ and $\cN=3$ vacua. Moreover, these critical points exhibit their full $\cN=2$ and $\cN=3$ supersymmetry within the $\cN=4$ $\textrm{SO}(3)_{\textrm{R}}$-invariant sector, as an explicit construction of the $\mathcal{N}=4$ gravitino mass matrix confirms. In section~\ref{sec:CanonicalN=4} we perform checks on our formalism. Firstly, the $\textrm{SO}(3)_{\textrm{R}}$-invariant sector is embedded, following \cite{Dibitetto:2011eu}, into the largest $\cN=4$ model contained within the $\cN=8$ ISO(7) theory. Secondly, the model is cast into the canonical $\cN=4$ embedding tensor formalism of \cite{Schon:2006kz}. This is interesting in its own right, as the resulting gauging is new from the point of view of half-maximal supergravity. In section~\ref{sec:further_trunc} the SU(3) and $\textrm{SO}(3)_{\textrm{d}} \times \textrm{SO}(3)_{\textrm{R}}$ invariant sectors of \cite{Guarino:2015qaa} are recovered. In section~\ref{sec:vacua} a thorough numerical scan for new $\textrm{SO}(3)_{\textrm{R}}$-invariant vacua is performed. Section~\ref{sec:Discussion} concludes and two appendices contain further details on group theory and the duality hierarchy.

\section{$\textrm{SO}(3)_{\textrm{R}}$-invariant sector of $\,\textrm{ISO}(7)\,$ maximal supergravity}
\label{sec:embedding_N8}

Our starting point is the dyonic $\,\textrm{ISO}(7)$-gauged maximal supergravity as presented in sec.~$2$ of \cite{Guarino:2015qaa} using the standard $\,\textrm{SL}(8)\,$ symplectic frame of $\,\mathcal{N}=8\,$ supergravity. We will perform a truncation of this theory to its $\,\textrm{SO}(3)_{\textrm{R}}$-invariant sector.

\subsection{Field content} \label{sec:fieldContent}

In order to determine the field content of the $\,\textrm{SO}(3)_{\textrm{R}}$-invariant sector of the dyonic $\,\textrm{ISO}(7)$-gauged maximal supergravity, one needs to know how the $\,\textrm{SO}(3)_{\textrm{R}}\,$ subgroup is embedded into $\,\textrm{ISO}(7)\,$. This group-theoretical embedding is given by
\begin{equation}
\label{Embedding_SO3_R}
\begin{array}{cccccccccc}
& & & & \textrm{G}_{2} & \supset & \textrm{SU}(3)      \\
\textrm{ISO}(7) & \supset & \textrm{SO}(7) & \supset & & & & \supset & \textrm{SO}(3)_{\textrm{R}} \sim \textrm{SU}(2)_{\textrm{R}} \\
& & & & \textrm{SO}(3)' \times \textrm{SO}(4) & \supset & \textrm{SO}(3)_{\textrm{d}} \times \textrm{SO}(3)_{\textrm{R}}
\end{array}
\end{equation}
where $\,\textrm{SO}(4) \sim \textrm{SO}(3)_{\textrm{L}} \times \textrm{SO}(3)_{\textrm{R}}\,$, and $\,\textrm{SO}(3)_{\textrm{d}}\,$ is the diagonal subgroup inside the product $\,\textrm{SO}(3)' \, \times \, \textrm{SO}(3)_{\textrm{L}}\,$. The eight gravitini of the maximal supergravity multiplet transforming in the $\,\bar{\textbf{8}}\,$ of the R-symmetry group $\,\textrm{SU}(8)\,$ decompose under $\,\textrm{SU}(2)_{\textrm{R}} \,$ as $\,\bar{\textbf{8}} \rightarrow 4 \times \textbf{1} + 2 \times \textbf{2}\,$. This decomposition features four singlets thus reflecting the $\,\cN=4\,$ structure of the $\,\textrm{SO}(3)_{\textrm{R}}$-invariant sector. The canonical $\,\mathcal{N}=4\,$ formulation of this sector will be presented in sec.~\ref{sec:SO(3)_R} following the general framework of \cite{Schon:2006kz}.

For the $\,\textbf{21} \,+\, \textbf{7}\,$ vector fields spanning the $\,\textrm{ISO}(7) = \textrm{SO}(7) \ltimes \mathbb{R}^{7}\,$ gauging of the maximal theory it will be instructive to look at the branching rules
\begin{equation}
\begin{array}{ccccccc}
\textrm{SO}(7) & \supset & \textrm{SO}(3)' \times  \textrm{SO}(3)_{\textrm{L}} \times \textrm{SO}(3)_{\textrm{R}} & \supset &  \textrm{SO}(3)_{\textrm{d}} \times \textrm{SO}(3)_{\textrm{R}} \\[2mm]
\textbf{21} &  & (\textbf{3},\textbf{1},\textbf{1}) + (\textbf{1},\textbf{3},\textbf{1}) + (\textbf{1},\textbf{1},\textbf{3}) + (\textbf{3},\textbf{2},\textbf{2})&  & 2 \times (\textbf{3},\textbf{1}) + (\textbf{1},\textbf{3}) + (\textbf{2}+\textbf{4} , \textbf{2})  \\[2mm]
\textbf{7} &  & (\textbf{3},\textbf{1},\textbf{1}) + (\textbf{1},\textbf{2},\textbf{2})  &  & (\textbf{3},\textbf{1}) + (\textbf{2},\textbf{2})  
\end{array}
\end{equation}
The electric and magnetic $\,\textrm{SO}(3)_{\textrm{R}}$-invariant vectors are denoted 
\begin{equation}
\label{A_fields_SO(3)_R}
A^{\Lambda}=(A'^{i} \, , \, A^{(\textrm{L})a} \, , \, A^{(\textrm{t})i})
\hspace{8mm} \textrm{ and } \hspace{8mm}  
\tilde{A}_{\Lambda}=(\tilde{A}'_{i} \, , \, \tilde{A}^{(\textrm{L})}_{a} \, , \, \tilde{A}^{(\textrm{t})}_{i}) \ ,
\end{equation}
with $\, i=1,2,3\,$ and $\,a=1,2,3\,$. The vectors $\,(A'^{i} \, , \, \tilde{A}'_{i}) \equiv (\textbf{3},\textbf{1}) \subset \textbf{21}\,$ and $\,(A^{(\textrm{t})i} \, , \, \tilde{A}^{(\textrm{t})}_{i}) \equiv (\textbf{3},\textbf{1})\subset \textbf{7}\,$ are associated with $\,\textrm{ISO}(3)' \subset \textrm{ISO}(7)\,$ whereas $\,(A^{(\textrm{L})a} \, , \, \tilde{A}^{(\textrm{L})}_{a})  \equiv (\textbf{3},\textbf{1})\subset \textbf{21}\,$ are associated with $\,\textrm{SO}(3)_{\textrm{L}} \subset \textrm{ISO}(7)\,$. Together they specify a truncated gauge group of the form
\begin{equation}
\label{G_trunc}
\textrm{G}=\textrm{ISO}(3)' \times \textrm{SO}(3)_{\textrm{L}} \ ,
\end{equation}
which, in $\,\cN=4\,$ terms, corresponds to the supergravity multiplet being coupled to three vector multiplets. More details about the group-theoretical embedding of the vector fields into maximal supergravity are presented in appendix~\ref{sec:N8_embedding}.

The $\,\textrm{SO}(3)_{\textrm{R}}$-invariant (pseudo-)scalars can be identified by performing the group-theoretical decomposition
\begin{equation}
\begin{array}{ccccccc}
\textrm{SO}(7) & \supset &  \textrm{SO}(3)_{\textrm{d}} \times \textrm{SO}(3)_{\textrm{R}} \\[2mm]
\textbf{1}  &  & (\textbf{1},\textbf{1}) \\[2mm]
\textbf{7} &  &  (\textbf{3},\textbf{1})  + (\textbf{2},\textbf{2})\\[2mm]
\textbf{27}  &  & (\textbf{5}+\textbf{1},\textbf{1}) +  (\textbf{3},\textbf{3}) +  (\textbf{4}+\textbf{2},\textbf{2})\\[2mm]
\textbf{35} \,\, (\textrm{pseudo-scalars}) &  & (\textbf{1},\textbf{1}) + (\textbf{5}+\textbf{3}+\textbf{1},\textbf{1}) + (\textbf{3},\textbf{3})  +  (\textbf{2},\textbf{2}) +  (\textbf{4}+\textbf{2},\textbf{2})
\end{array}
\end{equation}
The two fields denoted $\,\chi \equiv (\textbf{1},\textbf{1}) \subset \textbf{35}\,$ and $\,\varphi \equiv (\textbf{1},\textbf{1}) \subset \textbf{1}\,$ correspond to the (complex) axion-dilaton $\,\tau=\chi+i e^{-\varphi}\,$ of $\,\cN=4\,$ supergravity parameterising a scalar matrix
\begin{equation}
\label{M_alphabeta}
M_{\alpha \beta} = 
\frac{1}{\textrm{Im}\tau}
\left(
\begin{array}{cc}
1 & \chi \\
\chi & |\tau|^2
\end{array}
\right)
 \in \frac{\textrm{SL}(2)}{\textrm{SO}(2)} \ ,
\end{equation}
with $\,\alpha=+,-\,$ being a fundamental index of $\,\textrm{SL}(2)\,$. 
The rest of the scalars serve as coordinates in a coset space
\begin{equation}
M_{MN} = \mathcal{V}_{6,3}^{\textrm{T}} \, \mathcal{V}_{6,3} \in \frac{\textrm{SO}(6,3)}{\textrm{SO}(6) \times \textrm{SO}(3)} \ ,
\end{equation}
where $\,M, N\,$ are fundamental indices of $\,\textrm{SO}(6,3)\,$. These scalars are assembled into $\,3 \times 3\,$ matrices $\,\bm \nu \equiv   (\textbf{5}+\textbf{1},\textbf{1}) \subset \textbf{27} \,$, $\,\bm a \equiv   (\textbf{3},\textbf{1}) \subset \textbf{7} \,$ and $\,\bm b \equiv   (\textbf{5}+\textbf{3}+\textbf{1},\textbf{1}) \subset \textbf{35}\,$ so that
\begin{equation}
\label{Coset_63_N8}
\begin{array}{lll}
\mathcal{V}_{6,3} &=& \left(
\begin{array}{ccc}
{\bm \nu}^{-\textrm{T}} & 0  & 0 \\
0 & \mathbb{I}_3 & 0  \\
0 & 0 &  {\bm \nu}
\end{array}
\right)
\left(
\begin{array}{ccc}
 \mathbb{I}_3 & 0  & \bm{a} \\
0 & \mathbb{I}_3 & 0  \\
0 & 0 &   \mathbb{I}_3
\end{array}
\right)
\left(
\begin{array}{ccc}
 \mathbb{I}_3 \; & \bm{b}^{\textrm{T}}  \; & \tfrac12 \, \bm{b}^{\textrm{T}}  \bm{b} \\
0 & \mathbb{I}_3 & \bm{b}  \\
0 & 0 &   \mathbb{I}_3
\end{array}
\right) \ .
\end{array}
\end{equation}
The matrix $\,\bm \nu\,$ is in turn a coset element $\,\bm \nu \in \textrm{GL}(3)'/\textrm{SO}(3)\,$ which can be parameterised using scalars $\,\phi_i\,$ and $\,h^i{}_j\,$ as
\begin{eqnarray}
\label{GL3coset}
{\bm \nu} \equiv 
\left(
\begin{array}{ccc}
e^{-\frac{1}{\sqrt{2}} \phi_1} & e^{-\frac{1}{\sqrt{2}} \phi_1} \, h^1{}_2  & e^{-\frac{1}{\sqrt{2}} \phi_1}  \big(h^1{}_3 +\tfrac12 h^1{}_2 \, h^2{}_3    \big) \\[5pt]
0 & e^{-\frac{1}{\sqrt{2}} \phi_2} & e^{-\frac{1}{\sqrt{2}} \phi_2}  \, h^2{}_3   \\[5pt]
0 & 0 & e^{-\frac{1}{\sqrt{2}} \phi_3}
\end{array}
\right) \ ,
\end{eqnarray} 
and determine a scalar matrix $\,{\bm m} \equiv  {\bm \nu}^\textrm{T} {\bm \nu}\,$. The matrices $\,\bm{m}\,$, $\,\bm{a}\,$ and $\,\bm{b}\,$ have components $\,m_{ij} = m_{ji}\,$, $\,a_{ij} = - a_{ji}\,$ and $\,b^{a}{}_{j}\,$ in order to fit the appropriate representations. More details about the group-theoretical embedding of the (pseudo-) scalars into maximal supergravity are presented in appendix~\ref{sec:N8_embedding}. 

The $\,\textbf{21} + \textbf{27} + \textbf{7}\,$ two-form fields entering the restricted tensor hierarchy in the $\,\textrm{ISO}(7)\,$ maximal supergravity have a group-theoretical decomposition of the form
\begin{equation}
\scalebox{0.96}{$
\begin{array}{ccccccc}
\textrm{SO}(7) & \supset & \textrm{SO}(3)' \times  \textrm{SO}(3)_{\textrm{L}} \times \textrm{SO}(3)_{\textrm{R}} & \supset &  \textrm{SO}(3)_{\textrm{d}} \times \textrm{SO}(3)_{\textrm{R}} \\[2mm]
\textbf{21} &  & (\textbf{3},\textbf{1},\textbf{1}) + (\textbf{1},\textbf{3},\textbf{1}) + (\textbf{1},\textbf{1},\textbf{3}) + (\textbf{3},\textbf{2},\textbf{2})&  & 2 \times (\textbf{3},\textbf{1}) + (\textbf{1},\textbf{3}) + (\textbf{2}+\textbf{4} , \textbf{2})  \\[2mm]
\textbf{27} &  & (\textbf{5},\textbf{1},\textbf{1}) + (\textbf{1},\textbf{1},\textbf{1}) + (\textbf{1},\textbf{3},\textbf{3}) + (\textbf{3},\textbf{2},\textbf{2})&  &  (\textbf{5},\textbf{1}) + (\textbf{1},\textbf{1}) + (\textbf{3},\textbf{3}) + (\textbf{2} + \textbf{4} ,\textbf{2})  \\[2mm]
\textbf{7} &  & (\textbf{3},\textbf{1},\textbf{1}) + (\textbf{1},\textbf{2},\textbf{2})  &  & (\textbf{3},\textbf{1}) + (\textbf{2},\textbf{2})  
\end{array}
$}
\end{equation}
which implies a total of fifteen two-forms in the $\,\textrm{SO}(3)_{\textrm{R}}$-invariant sector. More concretely they sit in the representations
\begin{equation}
\label{two-form_fields}
\begin{array}{llll}
{B^{\scaleto{(A)}{6pt}}}_{i}{}^{j} \equiv (\textbf{3},\textbf{1}) \subset \textbf{21}
&\hspace{4mm} , & \hspace{4mm}
B_{a} \equiv (\textbf{3},\textbf{1}) \subset \textbf{21} & , \\[2mm]
{B^{\scaleto{(S)}{6pt}}}_{i}{}^{j} \equiv (\textbf{5},\textbf{1}) + (\textbf{1},\textbf{1})  \subset \textbf{27}
&\hspace{4mm} , & \hspace{4mm}
B^{i} \equiv (\textbf{3},\textbf{1}) \subset \textbf{7} & .
\end{array}
\end{equation}
Only the tensor fields $\,B^{i}\,$ enter the Lagrangian of the $\,\textrm{SO}(3)_{\textrm{R}}$-invariant sector. However, and despite not carrying independent dynamics, all the tensor fields in (\ref{two-form_fields}) enter the truncated tensor hierarchy as discussed in appendix~\ref{sec:TensHier}.

Lastly, the group-theoretical decomposition of the $\,\textbf{1} + \textbf{27}\,$ three-form potentials dual to electric components of the embedding tensor in the $\,\textrm{ISO}(7)\,$ maximal theory is given by
\begin{equation}
\scalebox{0.96}{$
\begin{array}{ccccccc}
\textrm{SO}(7) & \supset & \textrm{SO}(3)' \times  \textrm{SO}(3)_{\textrm{L}} \times \textrm{SO}(3)_{\textrm{R}} & \supset &  \textrm{SO}(3)_{\textrm{d}} \times \textrm{SO}(3)_{\textrm{R}} \\[2mm]
\textbf{1} &  & (\textbf{1},\textbf{1},\textbf{1}) &  & (\textbf{1},\textbf{1})   \\[2mm]
\textbf{27} &  & (\textbf{5},\textbf{1},\textbf{1}) + (\textbf{1},\textbf{1},\textbf{1}) + (\textbf{1},\textbf{3},\textbf{3}) + (\textbf{3},\textbf{2},\textbf{2})&  &  (\textbf{5},\textbf{1}) + (\textbf{1},\textbf{1}) + (\textbf{3},\textbf{3}) + (\textbf{2} + \textbf{4} ,\textbf{2})  \\[2mm]
\end{array}
$}
\end{equation}
thus yielding a total of seven three-forms in the $\,\textrm{SO}(3)_{\textrm{R}}$-invariant sector
\begin{equation}
\label{three-forms_electric}
C^{0} \equiv (\textbf{1},\textbf{1}) \subset \textbf{1}
\hspace{8mm} , \hspace{8mm}
C^{ij} \equiv (\textbf{5},\textbf{1}) + (\textbf{1},\textbf{1}) \subset \textbf{27} \ .
\end{equation}

\subsection{Lagrangian and equations of motion}

The bosonic Lagrangian of the $\,\textrm{SO}(3)_{\textrm{R}}$-invariant sector of $\,\textrm{ISO}(7)\,$ maximal supergravity can be obtained by direct truncation of the one in \cite{Guarino:2015qaa}. The result is given by
\begin{equation}
\label{L_SO(3)_R_N8}
\begin{array}{llll}
{\cal L} &=&   (R - V ) \,  * 1 + h_{uv} \,  D q^{u} \wedge * Dq^{v} + \tfrac12 \, {\cal I}_{\Lambda \Sigma} \, H_\2^\Lambda  \wedge * H_\2^\Sigma + \tfrac12 \, {\cal R}_{\Lambda \Sigma} \, H_\2^\Lambda  \wedge  H_\2^\Sigma \\[2mm]
&& + \, \tfrac14 \, m \, \epsilon^{ij}{}_k \, \tilde{A}^{(\textrm{t})}_i \wedge \tilde{A}^{(\textrm{t})}_j \wedge dA^{\prime k}  +\tfrac18 \, g \, m \, \tilde{A}^{(\textrm{t})}_i \wedge \tilde{A}^{(\textrm{t})}_j \wedge A^{\prime i} \wedge A^{\prime j}  \\[2mm]
& & + \, \tfrac12 \, g \, m \, \delta_{ij} B^i \wedge B^j  - m \, B^i \wedge \tilde{H}^{(\textrm{t})}_{\2 i}   \ ,
\end{array}
\end{equation}
where $\,g\,$ is the gauge coupling constant and $\,m\,$ is the magnetic charge in the theory. We have denoted the scalar kinetic terms collectively as
\begin{equation}
h_{uv} \,  D q^{u} \wedge * Dq^{v} = - \tfrac{1}{4} \, D M_{\alpha \beta} \wedge * DM^{\alpha \beta} - \tfrac{1}{8} \, D M_{MN}  \wedge *   D M^{MN} \ .
\end{equation}
The kinetic terms for the scalars in the $\,\textrm{SL}(2)/\textrm{SO}(2)\,$ coset read
\begin{equation}
\label{ScalKinTermsSL2}
-\tfrac{1}{4} \, D M_{\alpha \beta} \wedge * DM^{\alpha \beta} = \tfrac{1}{2} \, (d\varphi)^2 +  \tfrac12 \,  e^{2 \varphi} \, (d\chi )^2 \ ,
\end{equation}
where $\,(d\varphi)^2 \equiv d\varphi \wedge *  d\varphi\,$, etc. Those for the scalars in the $\,\textrm{SO}(6,3)/(\textrm{SO}(6) \times \textrm{SO}(3))\,$ coset are given by
\begin{eqnarray}
\label{ScalKinTerms63}
 - \tfrac{1}{8} \, D M_{MN}  \wedge *   D M^{MN}  &=&   -\tfrac14 \, \textrm{tr} \, \big( D  \bm{m} \, D  \bm{m}^{-1}   \big) +\tfrac12 \, \textrm{tr} \, \big(  \bm{m}^{-1} \, D \bm{b}^\textrm{T} \, D\bm{b} \big) \\[2mm]
&&+ \tfrac14 \, e^{ \sqrt{2} ( \phi_1 +\phi_2 +\phi_3 ) } \Big( 2 \,  \textrm{tr} \, \big( \bm{m} \bm{f} \bm{f}  \big)  - \textrm{tr} ( \bm{m} ) \, \textrm{tr} \, (  \bm{f} \bm{f} ) \Big) \ , \nonumber
\end{eqnarray}
where the matrix $\,\bm{f}\,$ has components $\,f_{ij}\,$ of the form
\begin{equation} 
\label{eq:fij=Daij}
f_{ij} = D a_{ij} +\delta_{ab} \, b^{a}{}_{[i} \, D b^{b}{}_{j]} \ .
\end{equation}

As usual in supergravity theories, the scalar fields couple both minimally and non-minimally to the vectors. The minimal couplings are governed by covariant derivatives of the form
\begin{equation}
\label{CovDers}
\begin{array}{lll}
D m _{ij} &=& d m _{ij} + 2 \,g \, A^{\prime h} \, \epsilon_{h(i}{}^k \,  m_{j) k }  \ ,  \\[2mm]
D a_{ij} &=& d a_{ij} + 2 \, g \,  A^{\prime h} \, \epsilon_{h [j }{}^k    a_{i] k }  + \epsilon_{kij} \, \big( \,  g \,  A^{(\textrm{t}) k } - m \, \delta^{kh} \tilde{A}^{(\textrm{t}) }_h \,   \big) \ , \\[2mm]
D b^{a}{}_i &=& d b^{a}{}_i + g \, A^{\prime h}  \, \epsilon_{hi}{}^k  \,   b^{a}{}_k - g \,  A^{(\textrm{L})  c } \,  \epsilon_{ c b }{}^{a} \, b^{ b }{}_i  \ .
\end{array}
\end{equation}
The non-minimal couplings occur via the scalar-dependent gauge kinetic matrix in (\ref{L_SO(3)_R_N8}) 
\begin{equation}
\label{N_kin_matrix}
\cN_{\Lambda \Sigma} = {\cal R}_{\Lambda \Sigma} + i \, {\cal I}_{\Lambda \Sigma} \ , 
\end{equation}
and involve the electric field strengths 
\begin{equation}
\label{N=4electricFS}
\begin{array}{lll}
H_\2^{\prime  i } &=& d A^{\prime  i } + \tfrac12 \, g \, \epsilon^i{}_{jk} \, A^{\prime  j } \wedge A^{\prime  k } \ , \\[2mm]
H_\2^{(\textrm{L}) a } &=& d A^{ (\textrm{L}) a  } + \tfrac12  \, g \, \epsilon^{a}{}_{ b   c  } \, A^{ (\textrm{L})   b} \wedge A^{(\textrm{L})   c  } \ ,  \\[2mm]
H_\2^{( \textrm{t})  i } &=& d A^{( \textrm{t})  i  } +  g \, \epsilon^i{}_{jk} \, A^{\prime  j } \wedge A^{( \textrm{t})  k } - \tfrac12  \, m \, \epsilon^i{}_j{}^k   \, A^{\prime  j } \wedge \tilde{A}^{( \textrm{t})}_k + m \, B^i   \ .
\end{array}
\end{equation} 
Note the presence of the tensor fields $\,B^{i}\,$ in the field strengths $\,H_\2^{( \textrm{t})  i }\,$ as a consequence of the magnetic charge $\,m\,$ in the theory \cite{deWit:2005ub}. In the basis of (\ref{A_fields_SO(3)_R}), the matrix $\,\cN_{\Lambda \Sigma}\,$ in (\ref{N_kin_matrix}) takes the form
\begin{eqnarray}
\label{GaugeKinMat}
\cN = \cN^{\textrm{T}}  = 
\left(
\begin{array}{lll}
\cN_1 & \cN_2  & \cN_3 \\
\cN_2^{\textrm{T}} & \cN_4  & \cN_5   \\
\cN_3^{\textrm{T}} & \cN_5^{\textrm{T}}  & \cN_6
\end{array}
\right) \; , \qquad
\end{eqnarray} 
in terms of the following $\,3\times 3\,$ blocks
\begin{eqnarray} 
\label{GaugeKinMatBlocks}
\cN_1 &=& -i e^{\varphi}  \bm{m}  - \big( i e^{\varphi}  \chi \,  \bm{m} -\tfrac12  {\bm b}^\textrm{T} {\bm b}  - \bm{a} \big)   \bm{N}^{-1}  \big( i e^{\varphi}  \chi \,  \bm{m} -\tfrac12  {\bm b}^\textrm{T} {\bm b}  + \bm{a} \big)  \; , \nonumber \\[5pt]
\cN_2 &=&  \tfrac{1}{\sqrt{2}} \,  {\bm b}^\textrm{T} -  \tfrac{1}{\sqrt{2}} \,  ( - \chi +  i e^{-\varphi}  )  \big(  i e^{\varphi}  \chi \,  \bm{m} -\tfrac12  {\bm b}^\textrm{T} {\bm b}  - \bm{a} \big)   \bm{N}^{-1}  {\bm b}^\textrm{T}    \; , \nonumber \\[5pt]
\cN_3 &=& \big( i e^{\varphi}  \chi \,  \bm{m} -\tfrac12  {\bm b}^\textrm{T} {\bm b}  - \bm{a} \big)   \bm{N}^{-1}    \; , \nonumber \\[5pt]
\cN_4 &=& - \tfrac12 ( - \chi +  i e^{-\varphi}  ) \, \mathbb{I}_3 - \tfrac12( - \chi +  i e^{-\varphi}  )^2   \, {\bm b}  \bm{N}^{-1}  {\bm b}^\textrm{T}     \; , \nonumber \\[5pt]
\cN_5 &=&  \tfrac{1}{\sqrt{2}} \,  ( - \chi +  i e^{-\varphi}  )  \, {\bm b}  \bm{N}^{-1}     \; , \nonumber \\[5pt]
\cN_6 &=& - \bm{N}^{-1}     \ ,
\end{eqnarray}
with $\,\bm{N}\,$ being the matrix
\begin{equation}
\bm{N} \equiv -i e^{-\varphi} (1+ e^{2\varphi} \chi^2 ) \bm{m}  -( - \chi +  i e^{-\varphi}  )  \, {\bm b}^\textrm{T} {\bm b} \; .
\end{equation}

In $\,\mathcal{N}=4\,$ terms, the gauging of (\ref{G_trunc}) is turned on in the vector multiplet sector only, as can be seen from the covariant derivative $\,D\tau = d\tau\,$. In particular, the scalars $\,a_{ij}\,$ are charged under the $\,\textrm{SO}(3)'\,$ factor and correspond to St\"uckelberg fields for the $\,\mathbb{R}^3\,$ shifts gauged by the electric gauge fields $\,A^{(\textrm{t}) i }\,$ and their magnetic duals $\,\tilde{A}^{(\textrm{t})}_i\,$. Accordingly, the scalars $\,a_{ij}\,$ do not enter the scalar potential $\,V\,$ of the theory which reads
{\setlength\arraycolsep{2pt}
\begin{eqnarray} 
\label{V_SO3}
V &=& g^2 \Big[ -4 \, e^{ \varphi}  - 4 \, e^{\frac{1}{\sqrt{2}} ( \phi_1+  \phi_2+  \phi_3)} \,  \tr \, \big(  {\bm m} + \tfrac12 \, {\bm b}^\textrm{T} {\bm b}  \big)    \nonumber \\[2mm]
&& \quad \;   +\tfrac12 \, e^{-\varphi +\sqrt{2} ( \phi_1+  \phi_2+  \phi_3)} (1+ e^{2\varphi} \chi^2 ) \Big( 2 \,  \tr \, \big(   {\bm m} {\bm m} + \tfrac12 \, {\bm b}^\textrm{T} {\bm b}   {\bm m}  \big)  -  ( \tr \, {\bm m} )^2  + \tfrac14  \big( \tr \, ( {\bm b}^\textrm{T} {\bm b}  ) \big)^2  \Big)    \nonumber \\[2mm]
&&   \quad \; +\tfrac12 e^\varphi  \big( \tr \, ( {\bm b}^\textrm{T} {\bm b}  \,  {\bm m}^{-1} ) \big)^2 - \tfrac12 e^\varphi \,  \tr \, ( {\bm b}^\textrm{T} {\bm b} \,  {\bm m}^{-1}  {\bm b}^\textrm{T} {\bm b}  \, {\bm m}^{-1}  )  +  e^{\varphi +\sqrt{2} ( \phi_1+  \phi_2+  \phi_3)} (\det  {\bm b} )^2   \nonumber \\[2mm]
&& \quad  \; -\sqrt{2} \, \chi \, e^{\varphi +\sqrt{2} ( \phi_1+  \phi_2+  \phi_3)} ( \det  {\bm b} ) \,   \tr \, \big(  {\bm m} + \tfrac12 \, {\bm b}^\textrm{T} {\bm b}  \big)    \Big] \nonumber \\[2mm]
&&
+ \,  g \, m \, e^{\varphi +\sqrt{2} ( \phi_1+  \phi_2+  \phi_3)}   \Big( \sqrt{2} \,  \det  {\bm b} -\tfrac12 \chi \,  \tr \, ( {\bm b}^\textrm{T} {\bm b} )   \Big) + \tfrac12 \, m^2 \, e^{\varphi +\sqrt{2} ( \phi_1+  \phi_2+  \phi_3)}  \  .
\end{eqnarray}
}As we will show in sec.~\ref{sec:further_trunc}, the scalar potential in (\ref{V_SO3}) contains all the known AdS$_{4}$ vacua for the $\,\textrm{ISO}(7)\,$ maximal supergravity up to date, and five new non-supersymmetric AdS$_{4}$ vacua.

\subsubsection*{Equations of motion}

It is useful to write the equations of motion that derive from the $\,\textrm{SO}(3)_{\textrm{R}}$-invariant Lagrangian in (\ref{L_SO(3)_R_N8}). The variations under the scalars, the electric vectors, and the metric, yield
\begin{equation}
\label{EomsMain}
\begin{array}{rcll}
D \big( h_{uv} * D q^v \big)  -\tfrac12 ( \partial_u h_{vw} ) D q^v \wedge * Dq^w  +\tfrac{1}{2} \partial_u V \, \textrm{vol}_4 & & \\[2mm]
- \tfrac14 (  \partial_u {\cal I}_{\Lambda \Sigma} \big) \, H_\2^\Lambda  \wedge * H_\2^\Sigma - \tfrac14 (  \partial_u {\cal R}_{\Lambda \Sigma} \big) \, H_\2^\Lambda  \wedge  H_\2^\Sigma &=& 0  & ,\\[4mm]
D  \tilde{H}_{\2 \Lambda }  + 2 h_{uv} \, k_\Lambda^u * Dq^v &=& 0 & ,  \\[4mm]
h_{uv} D_\mu q^u D_\nu q^v + \tfrac{1}{2}  V g_{\mu\nu} 
-\tfrac12 {\cal I}_{\Lambda \Sigma} \big( H^\Lambda_{\mu\lambda} H^\Sigma_{\nu}{}^\lambda - \tfrac14 g_{\mu\nu}  H^\Lambda_{\rho \sigma} H^{\Sigma \, \rho \sigma} \big) & = & R_{\mu \nu} & .
\end{array}
\end{equation} 
In the electric-vector equation, $\,\tilde{H}_{\2 \Lambda}\,$ stands for the magnetic field strength,
\begin{equation} 
\label{VectorDualityRelations}
\tilde{H}_{\2 \Lambda } = {\cal R}_{\Lambda \Sigma} \, H^\Sigma_\2 + {\cal I}_{\Lambda \Sigma} \, * H^\Sigma_\2 \ ,
\end{equation}
and $\,k_\Lambda\,$ are the $\,\textrm{SO}(6,3)\,$ isometries along which the gauging is turned on. These can be read off from the covariant derivatives (\ref{CovDers}).

Two other fields enter the Lagrangian (\ref{L_SO(3)_R_N8}): the magnetic vectors $\,\tilde{A}^{(\textrm{t})}_{i}\,$ and the two-forms $\,B^i\,$. These carry dynamical degrees of freedom but not independent ones, in the sense that their equations of motion give duality equations that relate them to the scalars, the electric vectors and the metric. Indeed, the variation of (\ref{L_SO(3)_R_N8}) with respect to $\,\tilde{A}^{(\textrm{t})}_{i}\,$ reproduces the last three components, in the basis (\ref{A_fields_SO(3)_R}), of the vector-vector duality relations (\ref{VectorDualityRelations}). The variation under $\,B^i\,$ gives the duality relation
\begin{equation} 
\label{3FormDualityRelations}
H^i_{\3 } = h_{uv} \, k^{u i} * Dq^v  \ ,
\end{equation}
with $\,k^i = \epsilon^i{}_{jk} \, \partial_{ a_{jk}}\,$. More generally, dualisation conditions can be written for all the fields in the tensor hierarchy introduced in section \ref{sec:fieldContent}. Please refer to appendix~\ref{sec:TensHier} for the details.

\section{$\mathcal{N}=4\,$  canonical formulation: new gaugings of half-maximal supergravity} \label{sec:CanonicalN=4}

In this section we investigate two different half-maximal truncations of the ISO(7) maximal supergravity giving rise to new gaugings of half-maximal supergravity. We present them in a canonical $\,\mathcal{N}=4\,$ fashion using the embedding tensor formalism in the standard symplectic frame of $\,\mathcal{N}=4\,$ gauged supergravity \cite{Schon:2006kz}. The first truncation is obtained by modding out the ISO(7) theory by a discrete $\,\mathbb{Z}_{2}\,$ element along the lines of \cite{Dibitetto:2011eu}. The second truncation recovers the $\,\textrm{SO}(3)_{\textrm{R}}$-invariant sector presented in the previous section but in $\,\mathcal{N}=4\,$ language.

\subsection{Brief review of $\,\cN=4\,$ supergravity with $\,\textrm{G}\subset \textrm{SO}(6,n_{v})\,$} 
\label{sec:N=4sec}

We start with a quick review of the gauged $\,\cN=4\,$ supergravities in four dimensions constructed in \cite{Schon:2006kz}\footnote{We parameterise the $\,\textrm{SL}(2)\,$ matrix $\,M_{\alpha \beta}\,$ as in (\ref{M_alphabeta}) which slightly differs from the conventions adopted in \cite{Schon:2006kz}.}. The field content of these theories consists of the supergravity multiplet coupled to an arbitrary number $\,n_{v} \in \mathbb{N}\,$ of vector multiplets. 

The supergravity multiplet contains, as bosonic degrees of freedom, the metric $\,g_{\mu\nu}\,$, six vector fields $\,A_{\mu}{}^{\tilde{m}}\,$ with $\,\tilde{m}=1,...,6\,$ and a complex scalar $\,\tau=\chi + i e^{-\varphi}\,$ parameterising a scalar matrix
\begin{equation}
\label{M_alphabeta_N4}
M_{\alpha \beta} = 
\textrm{Re} \left( \mathcal{V}_{\alpha} \, \mathcal{V}^{*}_{\beta} \right)
\in \frac{\textrm{SL}(2)}{\textrm{SO}(2)}
\hspace{8mm} \textrm{ with } \hspace{8mm}
\mathcal{V}_{\alpha} = - \frac{1}{\sqrt{\textrm{Im}\tau}} \,\, ( 1 \, , \, \tau ) \ ,
\end{equation}
where $\,\alpha=+,-\,$ is a fundamental $\,\textrm{SL}(2)\,$ index. The scalar matrix in (\ref{M_alphabeta_N4}) can in turn be constructed from a coset representative $\,\mathcal{V}_{2}\,$ (see (\ref{V2_N=4})) as $\,M_{\alpha \beta}=\delta_{\underline{\alpha}\underline{\beta}} \, (\mathcal{V}_{2})^{\underline{\alpha}}{}_{\alpha} \, (\mathcal{V}_{2})^{\underline{\beta}}{}_{\beta}\,$, where the local index of the coset representative has been underlined.
The vector multiplets contain vector fields $\,A_{\mu}{}^{\tilde{a}}\,$ with $\,{\tilde{a}=1,...,n_{v}}\,$ and $\,6 n_{v}\,$ scalars parameterising a coset element 
\begin{equation}
\label{M_MN}
M_{MN} \in \frac{\textrm{SO}(6 , n_{v})}{\textrm{SO}(6) \times \textrm{SO}(n_{v})} \ ,
\end{equation} 
with $\,M=(\tilde{m},\tilde{a})\,$ being a fundamental index of $\,\textrm{SO}(6 , n_{v})$ in a time-like coordinate basis. The scalar matrix in (\ref{M_MN}) can also be constructed from a coset representative $\,\mathcal{V}_{6,n_{v}}\,$ as $\,M_{MN}=\delta_{\underline{M}\underline{N}} \, (\mathcal{V}_{6,n_{v}})^{\underline{M}}{}_{N} \, (\mathcal{V}_{6,n_{v}})^{\underline{N}}{}_{N}\,$, where again we have underlined the local index of the coset representative.

In its ungauged version the theory possesses a global $\,\textrm{SL}(2) \times \textrm{SO}(6,n_{v})\,$ symmetry, known as the duality group, which is a subgroup of the electromagnetic transformations $\,{\textrm{Sp}(12+2n_{v},\mathbb{R})}\,$. The electric vector fields and their magnetic duals transform in the fundamental representation of the duality group, namely $\,A_{\mu}{}^{\alpha M}\in (\textbf{2},\textbf{6}+\bm{n_{v}})\,$, and span an abelian $\,{\textrm{G}=\textrm{U}(1)^{6} \times \textrm{U}(1)^{n_{v}}}\,$ gauge symmetry under which the scalars are not charged. After performing a gauging of a (non-)abelian subgroup $\,\textrm{G} \subset \textrm{SL}(2) \times \textrm{SO}(6,n_{v})\,$, a gauged supergravity is obtained with the scalar fields being now charged under $\,\textrm{G}\,$. In addition, a non-trivial scalar potential $\,V\,$ is generated.

\subsubsection*{The Lagrangian}

The Lagrangian of the $\,\cN=4\,$ gauged supergravities becomes totally specified by a constant tensor $\,X_{\alpha M \beta N \gamma P}\,$ that determines the embedding of the gauge connection into the $\,{\textrm{Sp}(12+2n_{v},\mathbb{R})}\,$ group of electromagnetic transformations \cite{Schon:2006kz}. This embedding tensor takes the form 
\begin{equation}
\label{X_tensor}
X_{\alpha M \beta N \gamma P} = \Theta_{\alpha M}{}^{\mathcal{A}} \,\, [t_{\mathcal{A}}]_{ \beta N \gamma P} = -\epsilon_{\beta \gamma} \, f_{\alpha MNP} - \epsilon_{\beta \gamma} \, \eta_{M[N} \, \xi_{\alpha P]} - \epsilon_{\alpha (\beta}\, \xi_{\gamma) M} \, \eta_{NP} \ , 
\end{equation}
where the index $\,\mathcal{A}\,$ runs over the generators $\,[t_{\mathcal{A}}]_{ \beta N \gamma P}\,$ of $\,\textrm{G} \subset \textrm{SL}(2) \times \textrm{SO}(6,n_{v})\,$ that have been gauged and where $\,\eta_{MN}\,$ and $\,\epsilon_{\alpha \beta}\,$ are the bilinear invariant tensors of $\,\textrm{SO}(6,n_{v})\,$ and $\,\textrm{SL}(2)\,$ respectively\footnote{We use conventions where $\,\epsilon_{+-}=\epsilon^{+-}=1\,$ so that $\,\epsilon_{\alpha \beta} \, \epsilon^{\alpha \gamma} =\delta_{\beta}^{\gamma}\,$. The generators of the $\,{\textrm{SL}(2) \times \textrm{SO}(6,n_{v})}\,$ duality group take the form
\begin{equation}
[t_{\alpha M \beta N}]_{\gamma P \delta Q} = \epsilon_{\alpha \beta} \, \epsilon_{\gamma \delta} \, [t_{MN}]_{PQ} + \eta_{MN} \, \eta_{PQ} \, [t_{\alpha \beta}]_{\gamma \delta} \ ,
\end{equation}
with $\,[t_{MN}]^{PQ}=\delta_{M}^{[P} \, \delta_{N}^{Q]}\,$ and $\,[t_{\alpha \beta}]^{\gamma \delta}=\delta_{\alpha}^{(\gamma} \, \delta_{\beta}^{\delta)}\,$ being the $\,\textrm{SO}(6,n_{v})\,$ and $\,\textrm{SL}(2)\,$ generators in the fundamental representation.
}.
The embedding tensor in (\ref{X_tensor}) contains two irreducible representations of the duality group given by 
\begin{equation}
f_{\alpha MNP} = f_{\alpha [MNP]} 
\hspace{10mm} \textrm{and} \hspace{10mm}
\xi_{\alpha M} \ .
\end{equation}
Consistency of the gauge algebra requires a set of quadratic constraints on the parameters $\,f_{\alpha MNP}\,$ and $\,\xi_{\alpha M}\,$ specifying the gaugings of the theory (see eq.(2.20) in \cite{Schon:2006kz}). 

In this work we will concentrate on a class of gaugings of the form $\,\textrm{G}\subset \textrm{SO}(6,n_{v})$. Keeping the $\,\textrm{SL}(2)\,$ factor of the duality group ungauged requires
\begin{equation}
\label{X_tensor_f}
\xi_{\alpha M}=0
\quad\quad \Rightarrow \quad\quad
X_{\alpha M \beta N \gamma P} = \Theta_{\alpha M}{}^{QR} \,\, [t_{QR}]_{ \beta N \gamma P} = -\epsilon_{\beta \gamma} \, f_{\alpha MNP} \ ,
\end{equation}
which drastically simplifies the Lagrangian constructed in \cite{Schon:2006kz}. The bosonic part consists of three pieces
\begin{equation}
\label{L_bos}
\mathcal{L}_{\textrm{bos}} = \mathcal{L}_{\textrm{kin}} - V + \mathcal{L}_{\textrm{top}} \ .
\end{equation}
The kinetic terms, together with the generalised theta-angle, for the various supergravity fields are given by
\begin{equation}
\label{Lkin}
\begin{array}{lll}
e^{-1} \, \mathcal{L}_{\textrm{kin}} &=&  R + \tfrac{1}{4} \, D_{\mu} M_{\alpha \beta} \,  D^{\mu} M^{\alpha \beta} + \tfrac{1}{8} \, D_{\mu} M_{MN} \,  D^{\mu} M^{MN}  \\[2mm]
&& - \tfrac{1}{4} \, \textrm{Im} \tau \, M_{MN} \, \mathcal{H}_{\mu \nu}{}^{+M} \, \mathcal{H}^{\mu \nu \, +N} - \tfrac{1}{8e} \, \textrm{Re} \tau \, \eta_{MN} \, \varepsilon^{\mu\nu\rho\sigma} \, \mathcal{H}_{\mu \nu}{}^{+M} \, \mathcal{H}_{\rho \sigma}{}^{+N} \ ,
\end{array}
\end{equation}
where $\,M_{MN}\,$ is positive definite. The electric vector field strengths and the scalar covariant derivatives take the form
\begin{equation}
\label{H_field_strengths}
\mathcal{H}_{\mu \nu}{}^{+M} = 2 \, \partial_{[\mu} A_{\nu]}{}^{+M} - \eta^{MM'} \, f_{\gamma M'NP} \, A_{[\mu}{}^{\gamma N} \, A_{\nu]}{}^{+P} +   \eta^{MM'} \, f_{- M'NP} \, B_{\mu\nu}{}^{NP} \ ,
\end{equation}
and
\begin{equation}
\label{cov_deriv}
\begin{array}{lll}
D_{\mu} M_{\alpha \beta} & = & \partial_{\mu} M_{\alpha \beta} \ , \\[2mm] 
D_{\mu} M_{MN} & = & \partial_{\mu} M_{MN} - 2 \, A_{\mu}{}^{\gamma P} \, f_{\gamma PQ(M} \, M_{N)Q'} \, \eta^{QQ'} \ .
\end{array}
\end{equation}
The presence of non-trivial magnetic charges $\,f_{-MNP}\,$ in the theory requires to introduce a set of auxiliary tensor fields $\,B_{\mu\nu}{}^{MN}=B_{\mu\nu}{}^{[MN]}\,$.\footnote{We define the tensor fields as $\,B_{\mu\nu}{}^{(\textrm{here})}=\tfrac{1}{2} B_{\mu\nu}{}^{\cite{Schon:2006kz}}\,$.} These tensor fields enter the vector field strengths in (\ref{H_field_strengths}) and come along with their own tensor gauge transformations \cite{deWit:2005ub}. The gauging procedure also generates a non-trivial scalar potential that is quadratic in the gauging parameters and takes the form
\begin{equation}
\label{V}
\begin{array}{lll}
V &=& \tfrac{1}{4} \, f_{\alpha MNP}  \, f_{\beta QRS} \,  M^{\alpha \beta} \, \left[ \tfrac{1}{3} \, M^{MQ} \, M^{NR} \, M^{PS} + \left( \tfrac{2}{3} \, \eta^{MQ} - M^{MQ} \right) \, \eta^{NR} \, \eta^{PS} \right] \\[2mm]
& & - \tfrac{1}{9} \, f_{\alpha MNP}  \, f_{\beta QRS} \,  \epsilon^{\alpha \beta} \, M^{MNPQRS} \ ,
\end{array}
\end{equation}
with
\begin{equation}
\label{M6_tensor}
M_{MNPQRS} \equiv \epsilon_{\underline{\tilde{m}\tilde{n}\tilde{p}\tilde{q}\tilde{r}\tilde{s}}} \, \mathcal{V}^{\underline{\tilde{m}}}{}_{M} \, \mathcal{V}^{\underline{\tilde{n}}}{}_{N} \, \mathcal{V}^{\underline{\tilde{p}}}{}_{P} \, \mathcal{V}^{\underline{\tilde{q}}}{}_{Q} \, \mathcal{V}^{\underline{\tilde{r}}}{}_{R} \, \mathcal{V}^{\underline{\tilde{s}}}{}_{S} \ .
\end{equation}
Local (underlined) indices in (\ref{M6_tensor}) are in a time-like coordinate basis for $\,\textrm{SO}(6,n_{v})$, namely, $\,\eta_{\textrm{time-like}}=\textrm{diag}(-\mathbb{I}_{6},\mathbb{I}_{n_{v}})\,$. These are related to light-like coordinates by the orthogonal transformation
\begin{equation}
\begin{array}{lllll}
i) & O &=&\left(
\begin{array}{ccc}
-\tfrac{1}{\sqrt{2}} \, \mathbb{I}_{n_{v}} & 0 & \tfrac{1}{\sqrt{2}} \, \mathbb{I}_{n_{v}} \\[2mm]
0 & -\mathbb{I}_{6-n_{v}}  & 0 \\[2mm]
\phantom{-}\tfrac{1}{\sqrt{2}} \, \mathbb{I}_{n_{v}} & 0 & \tfrac{1}{\sqrt{2}} \, \mathbb{I}_{n_{v}} 
\end{array}
\right)
&  \textrm{ for } \hspace{5mm} n_{v} \le 6 \ ,  \\[15mm]
ii) & O &=&\left(
\begin{array}{ccc}
-\tfrac{1}{\sqrt{2}} \, \mathbb{I}_{6} & \tfrac{1}{\sqrt{2}} \, \mathbb{I}_{6} & 0 \\[2mm]
\phantom{-}\tfrac{1}{\sqrt{2}} \, \mathbb{I}_{6} & \tfrac{1}{\sqrt{2}} \, \mathbb{I}_{6} & 0 \\[2mm]
0 & 0 & \mathbb{I}_{n_{v}-6}
\end{array}
\right)
&  \textrm{ for } \hspace{5mm} n_{v} \ge 6 \ ,
\end{array}
\end{equation}
so that $\,\eta_{\textrm{light-like}}=O \, \eta_{\textrm{time-like}}\, O^{t}\,$. Lastly, whenever there are non-trivial magnetic charges in the theory, there is a topological term involving the vector fields as well as the auxiliary tensor fields 
\begin{equation}
\label{Ltop}
\begin{array}{lll}
\mathcal{L}_{\textrm{top}} &=& -\tfrac{1}{2} \, \varepsilon^{\mu\nu\rho\lambda} \, \left[ -f_{-MNP} \, A_{\mu}{}^{-M} \, A_{\nu}{}^{+N} \, \partial_{\rho} A_{\lambda}{}^{-P} \right. \\[2mm]
& & \left. \quad \quad \quad \quad \,\,   - \tfrac{1}{4} \, f_{\alpha MNR} \, f_{\beta PQR'}\, \eta^{RR'} \, A_{\mu}{}^{\alpha M} \, A_{\nu}{}^{+N}  \, A_{\rho}{}^{\beta P}  \, A_{\lambda}{}^{-Q}  \right. \\[2mm]
&& \left. \quad \quad \quad \quad \,\, + \tfrac{1}{4} \, f_{+ MNP} \, f_{-M'QR} \, \eta^{MM'} \, B_{\mu\nu}{}^{NP} \, B_{\rho \lambda}{}^{QR} \right. \\[2mm]
&& \left. \quad \quad \quad \quad \,\,  - \tfrac{1}{2} \, f_{-MNP} \,  B_{\mu\nu}{}^{NP} \left( 2 \partial_{\rho} A_{\lambda}{}^{- M} - f_{\alpha QRM'}\, \eta^{MM'} \,  A_{\rho}{}^{\alpha Q} \, A_{\lambda}{}^{- R} \right)  \right] \ .
\end{array}
\end{equation}
This concludes our review of the main features of the $\,\cN=4\,$ gauged supergravities constructed in \cite{Schon:2006kz} when the gauging belongs to the class $\,\textrm{G}\subset \textrm{SO}(6,n_{v})\,$. In the following we will restrict to the case with $\,n_{v} \le 6\,$. This is a necessary condition for an $\,\cN=4\,$ (half-maximal) gauged supergravity to be embeddable into an $\,\cN=8\,$ (maximal) theory as the $\,\textrm{E}_{7(7)}\,$ duality group of the latter contains $\,\textrm{SL(2)} \times \textrm{SO}(6,6)\,$ as a maximal subgroup.

\subsection{Halving $\,\textrm{ISO}(7)\,$ maximal supergravity}

Let us consider the $\,\textrm{ISO}(7) \subset \textrm{E}_{7(7)}\,$ gauging of maximal supergravity put forward in \cite{Guarino:2015qaa}. This gauging is of dyonic type as a gauging of maximal supergravity\footnote{By dyonic here we mean dyonic in the standard $\,\textrm{SL}(8)\,$ symplectic frame of $\,\mathcal{N}=8\,$ supergravity, which differs from the one of $\,\mathcal{N}=4\,$ supergravity as we will see in a moment.}. From the general anaysis of \cite{Dibitetto:2011eu}, it can be consistently truncated to an $\,\cN=4\,$ gauging of the type discussed in the previous section with $\,n_{v}=6\,$ by means of a discrete (orientifold) $\,\mathbb{Z}_{2}\,$ projection\footnote{As discussed in sec.~$4.1$ of \cite{Inverso:2015viq}, the $\,\mathbb{Z}_{2}\,$ projection in (\ref{Z2-orientifold}) halving maximal supergravity in the standard $\,\textrm{SL}(8)\,$ symplectic frame yields an $\,\mathcal{N}=4\,$ Lagrangian which is not yet invariant under the full duality group of half-maximal supergravity. To achieve such an invariance one must perform a further dualisation of six vectors: three in the supergravity multiplet and three in the vector multiplets.}
\begin{equation}
\label{Z2-orientifold}
\mathbb{Z}_{2} \quad : \quad \textrm{E}_{7(7)} \rightarrow \textrm{SL(2)} \times \textrm{SO}(6,6) \ .
\end{equation}
The $\,\textrm{ISO}(7)\,$ gauging of the maximal theory consistently truncates to a gauging of the half-maximal theory. It is identified as
\begin{equation}
\label{G_nv=6}
\textrm{G}=\textrm{ISO}(3)' \times \textrm{SO}(4) = \textrm{ISO}(3)' \times \textrm{SO}(3)_{\textrm{L}}\times \textrm{SO}(3)_{\textrm{R}}  \ ,
\end{equation}
and its embedding into the $\,\textrm{SO}(6,6)\,$ factor of the duality group reads
\begin{equation}
\label{G-chain}
\textrm{ISO}(3)' \times \textrm{SO}(4) \subset \textrm{SO}(3,3)' \times  \textrm{SO}(3,3) \subset \textrm{SO}(6,6) \ , 
\end{equation}
with $\,\textrm{ISO}(3)' = \textrm{SO}(3)' \ltimes \mathbb{R}^3 \subset \textrm{SO}(3,3)'\,$ and $\,\textrm{SO}(4) \sim \textrm{SO}(3)_{\textrm{L}}\times \textrm{SO}(3)_{\textrm{R}} \subset \textrm{SO}(3,3)\,$.
Under the chain of embeddings in (\ref{G-chain}), the vector fields $\,A_{\mu}{}^{\alpha M} \in (\textbf{2},\textbf{12})\,$ in the theory decompose as
\begin{equation}
\label{vector-chain}
\begin{array}{ccccc}
\textrm{SL}(2) \times \textrm{SO}(6,6) & \supset & \textrm{SO}(1,1) \times \textrm{SO}(6,6) & \supset & \textrm{SO}(1,1) \times \textrm{SO}(3)'  \times \textrm{SO}(3)_{\textrm{L}}  \times \textrm{SO}(3)_{\textrm{R}} \\[2mm]
(\textbf{2},\textbf{12}) & &  \textbf{12}_{+} & & 2 \times (\textbf{3},\textbf{1},\textbf{1})_{+} +  (\textbf{1},\textbf{3},\textbf{1})_{+} +  (\textbf{1},\textbf{1},\textbf{3})_{+} \\[2mm]
 & & \textbf{12}_{-} & & 2 \times (\textbf{3},\textbf{1},\textbf{1})_{-} +  (\textbf{1},\textbf{3},\textbf{1})_{-} +  (\textbf{1},\textbf{1},\textbf{3})_{-}
\end{array}
\end{equation}
where the representations $\,2 \times (\textbf{3},\textbf{1},\textbf{1})_{\pm}\,$ correspond to electric ($+$) and magnetic ($-$) vector fields spanning the $\,\textrm{ISO}(3)'\,$ factor of $\,\textrm{G}\,$. The group-theoretical decomposition in (\ref{vector-chain}) translates into a splitting of the vector fields of the form
\begin{equation}
A_{\mu}{}^{\alpha M} =  
\left(
\begin{array}{c}
A_{\mu}{}^{+ M} \\[1mm]
A_{\mu}{}^{- M}
\end{array}
\right) = 
\left(
\begin{array}{c}
A_{\mu}{}^{M} \\[1mm]
\tilde{A}_{\mu}{}^{M}
\end{array}
\right) = 
\left(
\begin{array}{c}
A_{\mu}{}^{i} \\[1mm]
\tilde{A}_{\mu}{}^{i}
\end{array}
\right) 
\oplus
\left(
\begin{array}{c}
A_{\mu}{}^{a} \\[1mm]
\tilde{A}_{\mu}{}^{a}
\end{array}
\right) 
\oplus
\left(
\begin{array}{c}
A_{\mu}{}^{\bar{i}} \\[1mm]
\tilde{A}_{\mu}{}^{\bar{i}}
\end{array}
\right) 
\oplus
\left(
\begin{array}{c}
A_{\mu}{}^{\bar{a}} \\[1mm]
\tilde{A}_{\mu}{}^{\bar{a}}
\end{array}
\right)  \ ,
\end{equation}
where the index $\,M\,$ decomposes in a light-like coordinate basis as $\,{M=(i,a,\bar{i},\bar{a})}\,$ with $\,i=1,2,3\,$ and $\,a=1,2,3\,$. In this basis one has $\,\eta_{i\bar{j}}=\delta_{i\bar{j}}\,$ and $\,\eta_{a\bar{b}}=\delta_{a\bar{b}}\,$ so that
\begin{equation}
\eta_{MN} = \left(
\begin{array}{cc}
0 & \mathbb{I}_{6} \\[2mm]
\mathbb{I}_{6} &  0 
\end{array}
\right) \ .
\end{equation}

Gaugings of half-maximal supergravity of the form $\,\textrm{G} \subset \textrm{SO}(3,3)' \times  \textrm{SO}(3,3)\,$ have been extensively studied in the literature \cite{deRoo:1985jh,deRoo:2002jf,deRoo:2003rm,deRoo:2006ms,Roest:2009tt}, and further connected to non-geometric backgrounds in type II orientifold reductions \cite{Dibitetto:2011gm,Dibitetto:2012ia}. In these works, a non-trivial \mbox{de Roo-Wagemans} angle \cite{deRoo:1985jh} necessary to stabilise the SL(2) dilaton $\,\textrm{Im}\tau\,$ is generated by gauging one of the $\,\textrm{SO}(3,3)\,$  factors in (\ref{G-chain}) electrically and the other one magnetically, namely, $\,{\textrm{G}=\textrm{SO}(3,3)'_{+} \times \textrm{SO}(3,3)_{-}}\,$. However this is not the case for the half-maximal supergravity we are discussing here where the gauge factor inside $\,\textrm{SO}(3,3)'\,$ is gauged dyonically. This can be seen as follows.
Using light-like coordinates, the theory is specified in terms of an embedding tensor $\,f_{\alpha MNP}\,$ with non-vanishing components of the form \cite{Dibitetto:2011gm,Guarino:2015tja}
\begin{equation}
\label{ET_nv=6}
\begin{array}{rlllllllll}
\textrm{ISO}(3)' & : &   
f_{+\bar{i} \bar{j} k} &=& f_{+\bar{i} j \bar{k}} \,\,\, = \,\,\, f_{+i \bar{j} \bar{k}} \,\,\, = \,\,\,    g 
& , &
f_{-\bar{i}\bar{j}\bar{k}} &=& m  & ,
\\[2mm]
\textrm{SO}(4) & : & 
f_{-ab\bar{c}} &=& f_{-a\bar{b}c} \,\,\, = \,\,\, f_{-\bar{a}bc} \,\,\, = \,\,\,  - \sqrt{2} \, g 
& , & 
f_{-\bar{a}\bar{b}\bar{c}} &=&   - \sqrt{2} \, g 
& .
\end{array}
\end{equation}
It is worth noticing that the combination of vectors $\,g \, A_{\mu}{}^{i} + m \, \tilde{A}_{\mu}{}^{\bar{i}}\,$ serves as a gauge connection for $\,\mathbb{R}^{3} \subset \textrm{ISO}(3)' \subset \textrm{SO}(3,3)'\,$. In other words, an electric-magnetic deformation in maximal supergravity becomes the composition of an SL(2) (electric-magnetic) and a time/space-like deformation in half-maximal supergravity.

We will refrain from particularising the general Lagrangian of sec.~\ref{sec:N=4sec} to the case $\,n_{v}=6\,$ with an embedding tensor of the form (\ref{ET_nv=6}). Instead, we will perform a further truncation to a smaller $\,\textrm{SO}(3)_{\textrm{R}}$-invariant sector retaining only three vector multiplets.

\subsection{Truncation to the SO(3)$_{\textrm{R}}$-invariant sector}
\label{sec:SO(3)_R}

We perform a further truncation to the $\textrm{SO}(3)_{\textrm{R}}$-invariant sector of the theory by projecting out the three vector multiplets transforming as $\,(\textbf{1},\textbf{1},\textbf{3})_{\pm}\,$ in (\ref{vector-chain}). The resulting $\,\mathcal{N}=4\,$ theory therefore has $\,n_{v}=3\,$ and 
\begin{equation}
\label{G_nv=3}
\textrm{G} = \textrm{ISO}(3)' \times \textrm{SO}(3)_{\textrm{L}} \subset \textrm{SO}(6,3) \ ,
\end{equation}
in agreement with the gauge group in (\ref{G_trunc}). Using light-like coordinates $\,M=(i,a,\bar{i})\,$ for which $\,\eta_{i\bar{j}}=\delta_{i\bar{j}}\,$ and $\,\eta_{ab}=-\delta_{ab}\,$, namely,
\begin{equation}
\label{eta_SO(3)_R}
\eta_{MN} = \left(
\begin{array}{ccc}
0 & 0 & \mathbb{I}_{3} \\[2mm]
0 & -\mathbb{I}_{3} & 0 \\[2mm]
\mathbb{I}_{3} & 0 & 0
\end{array}
\right) \ ,
\end{equation}
the full $\,\cN=4\,$ Lagrangian is specified in terms of an embedding tensor $\,f_{\alpha MNP}\,$ with non-vanishing components of the form
\begin{equation}
\label{ET_SO(3)_R}
\begin{array}{rlllllllll}
\textrm{ISO}(3)' & : &   
f_{+\bar{i} \bar{j} k} &=& f_{+\bar{i} j \bar{k}} \,\,\, = \,\,\, f_{+i \bar{j} \bar{k}} \,\,\, = \,\,\,   g 
& , &
f_{-\bar{i}\bar{j}\bar{k}} &=&  m  & ,
\\[2mm]
\textrm{SO}(3)_{\textrm{L}} & : & f_{-abc} &=& - \sqrt{2} \, g  \,\,\,  .
\end{array}
\end{equation}
The gauge group $\,\textrm{G}= \textrm{ISO}(3)' \times \textrm{SO}(3)_{\textrm{L}}\,$ is associated with generators $\,[t_{MN}]^{P}{}_{Q}=2\,\delta_{[M}^{P}\eta_{N]Q}\,$ in $\,\textrm{SO}(6,3)\,$ given by
\begin{equation}
\label{GTL_generators}
G'_{i}= \epsilon_{i}{}^{jk} \, \left( \delta_{j}{}^{\bar{j}}\,  t_{\bar{j}k} + \delta_{k}{}^{\bar{k}} \, t_{j\bar{k}} \right)  
\hspace{5mm} , \hspace{5mm}
T'_{i} = \tfrac{1}{2} \, \epsilon_{i}{}^{jk} \, t_{jk}
\hspace{5mm} , \hspace{5mm}
L_{a} = \tfrac{1}{2} \, \epsilon_{a}{}^{bc} \, t_{bc} \ ,
\end{equation}
which satisfy commutation relations of the form 
\begin{equation}
[G'_{i},G'_{j}]=-\epsilon_{ij}{}^{k} \, G'_{k}
\hspace{5mm} , \hspace{5mm}
[G'_{i},T'_{j}]=-\epsilon_{ij}{}^{k} \, T'_{k}
\hspace{5mm} , \hspace{5mm}
[L_{a},L_{b}]=\epsilon_{ab}{}^{c} \, L_{c} \ .
\end{equation}
Combining the relation in (\ref{X_tensor_f}) with the embedding tensor components in (\ref{ET_SO(3)_R}) and the generators in (\ref{GTL_generators}) yields
\begin{equation}
\begin{array}{lcll}
\Theta_{+i}{}^{\,MN} \, t_{MN} = g \, T'_{i} &  ,   & \hspace{8mm} \Theta_{+\bar{i}}{}^{\,MN} \, t_{MN} = g \, \delta_{\bar{i}}{}^{i} \, G'_{i} & , \\[3mm]
\Theta_{-a}{}^{\,MN} \, t_{MN} = - \sqrt{2} \, g \, L_{a} & , & \hspace{8mm} \Theta_{-\bar{i}}{}^{\,MN} \, t_{MN} = m \, \delta_{\bar{i}}{}^{i} \, T'_{i} &  .
\end{array}
\end{equation}

The Lagrangian of this $\,\cN=4\,$ sector can be obtained from the general results of sec.~\ref{sec:N=4sec} by imposing $\,n_{v}=3\,$ and using the non-vanishing embedding tensor in (\ref{ET_SO(3)_R}). In this way we provide the canonical $\,\mathcal{N}=4\,$ formulation of the $\,\textrm{SO}(3)_{\textrm{R}}$-invariant sector presented in sec.~\ref{sec:embedding_N8}.

\subsection*{Lagrangian, duality relations and symplectic frames}

Together with the Einstein-Hilbert term, the Lagrangian of $\,\mathcal{N}=4\,$ supergravity consists on two pieces: a scalar sector and a vector-tensor sector. We discuss them separately here for the SO(3)$_{\textrm{R}}$-invariant sector of the $\,\textrm{ISO}(3)' \times \textrm{SO}(4)\,$ half-maximal supergravity.

\subsubsection*{\normalfont{$\circ$ \textit{Scalar sector}}}

There are $\,2 \,+\, 3 \times 6 = 20 \,$ scalar fields in the theory parameterising two coset elements 
\begin{equation}
\label{Vcosets}
\mathcal{V}_{2} \in \frac{\textrm{SL}(2)}{\textrm{SO}(2)}
\hspace{10mm} \textrm{ and } \hspace{10mm}
\mathcal{V}_{6,3} \in \frac{\textrm{SO}(6,3)}{\textrm{SO}(6) \times \textrm{SO}(3)} \ .
\end{equation}
The first one is obtained upon exponentiation of the Cartan generator $\,H\,$ and the positive root $\,E_{+}\,$ of the $\,\textrm{SL}(2)\,$ algebra, and takes the form
\begin{equation}
\label{V2_N=4}
\mathcal{V}_{2} = e^{\frac{1}{2} \, \varphi \, H} \, e^{\chi \, E_{+}} = 
\left(
\begin{array}{ll}
e^{\frac{1}{2}\varphi} & e^{\frac{1}{2}\varphi} \, \chi \\[2mm]
0 & e^{-\frac{1}{2}\varphi} 
\end{array}
\right) \ ,
\end{equation}
so that $\,M_{\alpha \beta}=\delta_{\underline{\alpha}\underline{\beta}} \, (\mathcal{V}_{2})^{\underline{\alpha}}{}_{\alpha} \, (\mathcal{V}_{2})^{\underline{\beta}}{}_{\beta}\,$ is given by (\ref{M_alphabeta}). The second one is obtained upon exponentiation of the Cartan generators $\,H^{i} = \sqrt{2} \, t_{i\bar{i}}\,$ and the positive roots $\,E_{i}{}^{j}=\delta_{i}{}^{\bar{i}} \, \delta^{jk} \, t_{\bar{i}k}\,$ ($\,i<j\,$), $\,V^{ij}=\delta^{ik} \, \delta^{jl} \, t_{kl}\,$ and $\,U_{a}{}^{j}=\delta^{jk} \, t_{ak}\,$ of the $\,\textrm{SO}(6,3)\,$ algebra in the light-like basis. It takes the form
\begin{equation}
\label{Coset_63_N4}
\begin{array}{lll}
\mathcal{V}_{6,3} &=& e^{\frac{1}{2} \, \phi_{i} \, H^{i}} \, e^{h^{i}{}_{j} \, E_{i}{}^{j}} \, e^{\frac{1}{2} \, a_{ij} \, V^{ij}} \, e^{b^{a}{}_{j} \, U_{a}{}^{j}}  \ ,
\end{array}
\end{equation}
and recovers the result in (\ref{Coset_63_N8}). The scalars $\,\phi_i\,$ and $\,h^i{}_j\,$ parameterise the coset element $\,\bm \nu \in \textrm{GL}(3)'/\textrm{SO}(3)\,$ as in (\ref{GL3coset}) and determine the scalar matrix $\,{\bm m} \equiv  {\bm \nu}^\textrm{T} {\bm \nu}\,$. The addition of the scalars $\,a_{ij}\,$ extends the coset $\,\textrm{GL}(3)'/\textrm{SO}(3)\,$ to $\,\textrm{SO}(3,3)'/\textrm{SO}(3)^2 \sim \textrm{SL}(4)'/\textrm{SO}(4)\,$, whereas adding the scalars $\,b^{a}{}_{j}\,$ extends the latter to the full $\,\mathcal{V}_{6,3}\,$ coset in (\ref{Vcosets}) with $\,M_{MN}=\delta_{\underline{M}\underline{N}} \, (\mathcal{V}_{6,3})^{\underline{M}}{}_{N} \, (\mathcal{V}_{6,3})^{\underline{N}}{}_{N}\,$. This symmetric matrix $\,M_{MN}\,$ has independent block components
\begin{equation}
\label{M_scalar_SO(3)R}
\begin{array}{lll}
M_{ij} & = & \bm m^{-1} \ ,  \\[2mm]
M_{ib} & = & \bm m^{-1} \, \bm b^{\textrm{T}} \ ,  \\[2mm]
M_{i\bar{j}} & = &  \bm m^{-1} \, \left( \tfrac{1}{2} \, \bm b^\textrm{T} \bm b  +  \bm a  \right) \ ,  \\[2mm]
M_{ab} & = & \mathbb{I}_{3} + \bm b \, \bm m^{-1} \, \bm b^{\textrm{T}}  \ ,  \\[2mm]
M_{a\bar{j}} & = & \bm b  + \bm b \, \bm m^{-1} \, \left( \tfrac{1}{2} \, \bm b^\textrm{T} \bm b  + \bm a  \right)  \ ,  \\[2mm]
M_{\bar{i}\bar{j}} & = &  \bm m + \bm b^\textrm{T} \bm b   + \left( \tfrac{1}{2} \, \bm b^\textrm{T} \bm b  -  \bm a  \right) \bm m^{-1}   \left( \tfrac{1}{2} \, \bm b^\textrm{T} \bm b  +  \bm a  \right)   \ .  
\end{array}
\end{equation}

Using differential form notation, the Einstein-Hilbert term and the scalar sector of the theory are given by 
\begin{equation}
\label{Einstein-scalar_SO(3)_R}
\begin{array}{lll}
\mathcal{L}_{\textrm{EH-s}} &=& (R -V ) \,  * 1 - \tfrac{1}{4} \, D M_{\alpha \beta} \wedge * DM^{\alpha \beta} - \tfrac{1}{8} \, D M_{MN}  \wedge *   D M^{MN} \ .
\end{array}
\end{equation}
The kinetic terms for the scalar fields are the same as in (\ref{ScalKinTermsSL2}) and (\ref{ScalKinTerms63}). The covariant derivatives for the various scalars can be extracted from (\ref{cov_deriv}) and read
\begin{equation}
\label{DM_SO(3)_R}
\begin{array}{llll}
Dm_{ij} & = & dm_{ij} + 2 \, g \, A^{\bar{k}} \, \epsilon_{\bar{k} (j}{}^{k} \, m_{i)k} & , \\[2mm]
Da_{ij} & = & da_{ij} + 2 \, g \, A^{\bar{k}} \, \epsilon_{\bar{k} [j}{}^{k} \, a_{i]k} - \big( \, g \, A^{k} \, \epsilon_{kij} + m \, \tilde{A}^{\bar{k}} \, \epsilon_{\bar{k}ij}  \, \big) & , \\[2mm]
Db^{a}{}_{i} & = & db^{a}{}_{i}  + g \, A^{\bar{k}} \, \epsilon_{\bar{k}i}{}^{k} \, b^{a}{}_{k} - \sqrt{2} \, g \, \tilde{A}^{c}\, \epsilon_{cb}{}^{a} \, b^{b}{}_{i}  & .
\end{array}
\end{equation}
Notice the combination of vectors $\,g \, A^{k} \, + m \, \tilde{A}^{\bar{k}} \,$ entering the gauge connection in $\,Da_{ij}\,$.

It is also worth emphasising here that the scalars $\,b^{a}{}_{j}\,$ are charged under both $\,\textrm{SO}(3,3)'\,$ and $\,\textrm{SO}(3)_{\textrm{L}} \subset \textrm{SO}(3,3)\,$ in (\ref{G-chain}), thus rendering this $\cN=4$ model different from the ones investigated in \cite{deRoo:1985jh,deRoo:2002jf,deRoo:2003rm,deRoo:2006ms,Roest:2009tt,Dibitetto:2011gm,Dibitetto:2012ia} where only scalars transforming separately under each of the $\textrm{SO}(3,3)$ factors were retained. Lastly, using the embedding tensor in (\ref{ET_SO(3)_R}) and the scalar matrix components in (\ref{M_scalar_SO(3)R}), we have verified that the scalar potential (\ref{V}) gives perfect agreement with the expression in (\ref{V_SO3}).

\subsubsection*{\normalfont{$\circ$ \textit{Vector-tensor sector}}}

Using differential form notation, the vector part of the Lagrangian (\ref{Lkin}) depends on the electric field strengths
\begin{equation}
\label{H2_elec}
\mathcal{H}_{\2}^{+M}=( \mathcal{H}_{\2}^{i} \, , \, \mathcal{H}_{\2}^{a} \, , \, \mathcal{H}_{\2}^{\bar{i}}) \ ,
\end{equation}
and reads
\begin{equation}
\label{Lkin_SO(3)_R}
\mathcal{L}_{\textrm{vec}}  =   
- \tfrac{1}{2} \, \textrm{Im} \tau \, M_{MN} \, \mathcal{H}_{\2}^{+M} \, \wedge * \mathcal{H}_{\2}^{+N}  - \tfrac{1}{2} \,  \textrm{Re} \tau \, \eta_{MN} \,  \mathcal{H}_{\2}^{+M} \wedge \mathcal{H}_{\2}^{+N} \ ,
\end{equation}
with scalar-dependent generalised gauge couplings given by (\ref{M_scalar_SO(3)R}) and field strengths of the form
\begin{equation}
\label{H_field_strengths_SO(3)_R}
\begin{array}{lll}
%
\mathcal{H}_{\2}^{i} &=& dA^{i} -  g \, \epsilon^{i}{}_{\bar{j}k} \, A^{\bar{j}} \wedge A^{k} - \tfrac{1}{2} \, m \, \epsilon^{i}{}_{\bar{j}\bar{k}} \, \tilde{A}^{\bar{j}} \wedge A^{\bar{k}}  + m \, \epsilon^{i}{}_{\bar{j}\bar{k}} \, B^{\bar{j}\bar{k}} \ , \\[2mm]
\mathcal{H}_{\2}^{a} &=& dA^{a} - \tfrac{1}{\sqrt{2}} \, g \, \epsilon^{a}{}_{bc} \, \tilde{A}^{b} \wedge A^{c} + \sqrt{2} \, g \, \epsilon^{a}{}_{bc} \, B^{bc} \ , \\[2mm]
\mathcal{H}_{\2}^{\bar{i}} &=& dA^{\bar{i}} - \tfrac{1}{2} \, g \, \epsilon^{\bar{i}}{}_{\bar{j}\bar{k}} \, A^{\bar{j}} \wedge A^{\bar{k}} \ .
\end{array}
\end{equation}
Note that the electric vectors $\,A^{a}\,$ are ungauged, namely, they do not enter the gauge connection in (\ref{DM_SO(3)_R}). Still they will propagate unlike the magnetic vectors ($\,\tilde{A}^{c} \, , \, \tilde{A}^{\bar{k}}\,$) and the tensor fields ($B^{ab} \, , \, \,B^{\bar{i}\bar{j}}\,$) which do not carry independent dynamics. Finally the topological term in (\ref{Ltop}) reduces to an expression of the form
%
%
\begin{equation}
\label{Ltop_SO(3)_R}
\begin{array}{lll}
\mathcal{L}_{\textrm{top}} &=&  - \tfrac{1}{\sqrt{2}} \, g \, \epsilon_{abc} \, \tilde{A}^{a} \wedge  A^{b} \wedge d\tilde{A}^{c} +  \tfrac{1}{2} \, m \, \epsilon_{\bar{i}\bar{j}\bar{k}} \, \tilde{A}^{\bar{i}} \wedge  A^{\bar{j}} \wedge d\tilde{A}^{\bar{k}}   \\[2mm]
&& + \, \tfrac{1}{8} \, g^2 \, \epsilon_{\bar{i}\bar{j}k} \, \epsilon_{\bar{m}n}{}^{k} \, ( A^{\bar{i}} \wedge  A^{\bar{j}}  \wedge  A^{\bar{m}}  \wedge \tilde{A}^{n}  +  A^{\bar{m}} \wedge  A^{n}  \wedge  A^{\bar{i}}  \wedge \tilde{A}^{\bar{j}} ) \\[2mm]
&& + \, \tfrac{1}{8} \, g^2 \, \epsilon_{\bar{i}\bar{j}k}\, \epsilon_{m\bar{n}}{}^{k} \, ( A^{\bar{i}} \wedge  A^{\bar{j}}  \wedge  A^{m}  \wedge \tilde{A}^{\bar{n}} + A^{m} \wedge  A^{\bar{n}}  \wedge  A^{\bar{i}}  \wedge \tilde{A}^{\bar{j}})  \\[2mm]
&& + \, \tfrac{1}{8} \, g \, m \, \epsilon_{\bar{i}\bar{j}k} \, \epsilon_{\bar{m}\bar{n}}{}^{k} ( A^{\bar{i}} \wedge  A^{\bar{j}}  \wedge  \tilde{A}^{\bar{m}}  \wedge \tilde{A}^{\bar{n}} +  \tilde{A}^{\bar{m}} \wedge  A^{\bar{n}}  \wedge  A^{\bar{i}}  \wedge \tilde{A}^{\bar{j}})  \\[2mm]
&& - \, \tfrac{1}{4} \, g^2 \,  \epsilon_{abe} \, \epsilon_{cd}{}^{e} \, \tilde{A}^{a} \wedge  A^{b}  \wedge  \tilde{A}^{c}  \wedge \tilde{A}^{d}   -  \tfrac{1}{2} \, g \, m \,  \epsilon^{\bar{i}}{}_{\bar{j}\bar{k}} \, \epsilon_{\bar{i}\bar{m}\bar{n}}  \, B^{\bar{j}\bar{k}} \wedge B^{\bar{m}\bar{n}}  \\[2mm]
&& - \, \sqrt{2} \, g \, \epsilon_{abc} \,  B^{bc} \wedge  \left(d\tilde{A}^{a} - \tfrac{1}{\sqrt{2}} \, g \,  \epsilon^{a}{}_{de}\,   \tilde{A}^{d} \wedge \tilde{A}^{e} \right)  \\[2mm]
&& + \, m \, \epsilon_{\bar{i}\bar{j}\bar{k}} \,  B^{\bar{j}\bar{k}} \wedge  \left(d\tilde{A}^{\bar{i}} - \tfrac{1}{2} \, g \,  \epsilon^{\bar{i}}{}_{\bar{m}\bar{n}}\,   A^{\bar{m}} \wedge \tilde{A}^{\bar{n}} \right)  \ .
\end{array}
\end{equation}
The above topological term (\ref{Ltop_SO(3)_R}) takes a lengthy form compared to the one in (\ref{L_SO(3)_R_N8}) due to the symplectic frame used in the $\,\cN=4\,$ Lagrangian of \cite{Schon:2006kz} and our choice of light-like coordinates for $\,\textrm{SO}(6,3)\,$. Note that the symplectic invariant matrix that results upon the $\,\textrm{SO}(3)_{\textrm{R}}$-invariant truncation of the $\,\cN=8\,$ theory is given by
\begin{equation}
\label{Omega_N8_truncation}
\begin{array}{ccc}
\Omega_{\textrm{Sp}(56,\mathbb{R})} = 
\left( \begin{array}{ll}
0 & \mathbb{I}_{28} \\[2mm]
-\mathbb{I}_{28} & 0
\end{array}  \right) & \longrightarrow & \Omega_{\textrm{Sp}(18,\mathbb{R})} = \left( \begin{array}{ll}
0 & \mathbb{I}_{9} \\[2mm]
-\mathbb{I}_{9} & 0
\end{array}  \right) \ ,
\end{array}
\end{equation}
whereas the canonical formulation of $\,\cN=4\,$ supergravity uses\footnote{The matrices $\,\Omega_{\textrm{Sp}(18,\mathbb{R})}\,$ and $\,\Omega'_{\textrm{Sp}(18,\mathbb{R})}\,$ in (\ref{Omega_N8_truncation}) and (\ref{Omega_N4}) are related by an $\,\textrm{SO}(18)\,$ rotation.} \cite{Schon:2006kz,Dibitetto:2011eu}
\begin{equation}
\label{Omega_N4}
\begin{array}{ccc}
\Omega'_{\textrm{Sp}(18,\mathbb{R})} = \epsilon_{\alpha \beta} \otimes \eta_{MN} 
=
\left( \begin{array}{ll}
0 & \eta \\[2mm]
-\eta & 0
\end{array}  \right) \ ,
\end{array}
\end{equation}
with $\,\eta\,$ given in (\ref{eta_SO(3)_R}). We will come back to this issue by the end of the section when discussing symplectic frames.

\subsubsection*{Duality relations}

Let us discuss the duality relations in the $\,\textrm{SO}(3)_{\textrm{R}}$-invariant sector. Similarly as for the electric vectors in (\ref{H2_elec}), one can introduce field strengths for the magnetic vectors
\begin{equation}
\label{H2_mag}
\mathcal{H}_{\2}^{-M}=( \tilde{\mathcal{H}}_{\2}^{i} \, , \, \tilde{\mathcal{H}}_{\2}^{a} \, , \, \tilde{\mathcal{H}}_{\2}^{\bar{i}}) \ .
\end{equation}
The magnetic vector field strengths will enter the equations of motion that follow from the $\,\textrm{SO}(3)_{\textrm{R}}$-invariant Lagrangian. More concretely, the equation of motion of the tensor fields yields a set of vector-vector duality relations between electric and magnetic field strengths
\begin{equation}
\label{Duality_relation_22_SO(3)_R}
\begin{array}{llll}
%
%
\tilde{\mathcal{H}}_{\2}{}^{i} &=&  - \,  \textrm{Im}\tau \,  \delta^{i\bar{i}} \, ( M_{\bar{i}k}  \, * \mathcal{H}_{\2}{}^{k} + M_{\bar{i}c}  \, * \mathcal{H}_{\2}{}^{c} + M_{\bar{i}\bar{k}}  \, * \mathcal{H}_{\2}{}^{\bar{k}}) + \textrm{Re}\tau \, \mathcal{H}_{\2}{}^{i} & , \\[2mm]
\tilde{\mathcal{H}}_{\2}{}^{a} &=& \phantom{-} \, \textrm{Im}\tau \,  \delta^{ab} \, ( M_{bk}  \, *  \mathcal{H}_{\2}{}^{k} + M_{bc}  \, *  \mathcal{H}_{\2}{}^{c} + M_{b\bar{k}}  \, *  \mathcal{H}_{\2}{}^{\bar{k}}) + \textrm{Re}\tau \, \mathcal{H}_{\2}{}^{a} & , \\[2mm]
\tilde{\mathcal{H}}_{\2}{}^{\bar{i}} &=&   - \,  \textrm{Im}\tau \,  \delta^{\bar{i}i} \, ( M_{ik}  \, * \mathcal{H}_{\2}{}^{k} + M_{ic}  \, * \mathcal{H}_{\2}{}^{c} + M_{i\bar{k}}  \, * \mathcal{H}_{\2}{}^{\bar{k}}) + \textrm{Re}\tau \, \mathcal{H}_{\2}{}^{\bar{i}} & , \\[2mm]
\end{array}
\end{equation}
with scalar-dependent matrices given in (\ref{M_scalar_SO(3)R}). On the other hand, the equation of motion of the magnetic vectors gives the tensor-scalar duality relations
\begin{equation}
\label{Duality_relation_31_SO(3)_R}
\begin{array}{llll}
%
\mathcal{H}_{\3}{}^{\bar{i}\bar{j}}  &=&\phantom{-} \delta^{\bar{i}i} \, ( M_{ik} \, * DM^{k\bar{j}} + M_{ic} \, * DM^{c\bar{j}} + M_{i\bar{k}} \, * DM^{\bar{k}\bar{j}} ) & , \\[2mm]
\mathcal{H}_{\3}{}^{ab} &=& - \delta^{ad}\, ( M_{dk} \, * DM^{kb} + M_{dc} \, * DM^{cb} + M_{d\bar{k}} \, * DM^{\bar{k}b} ) & .
\end{array}
\end{equation}
As a result of the duality relations in (\ref{Duality_relation_22_SO(3)_R}) and (\ref{Duality_relation_31_SO(3)_R}), the magnetic vectors and the tensor fields entering the $\,\textrm{SO}(3)_{\textrm{R}}$-invariant Lagrangian do not carry independent dynamics.

\subsubsection*{Symplectic frames}

Let us combine the generalised theta-angle and the gauge kinetic matrix for the electric vectors $\,A^{+M}\,$ in (\ref{Lkin_SO(3)_R}) into a matrix $\,\cN'\,$ of the form
\begin{equation}
\label{NkinSW}
\cN^\prime  = - \chi \, \eta  - i e^{-\varphi} M \ ,
\end{equation}
where $\,M_{MN}\,$ was built from the coset representative in (\ref{Coset_63_N4}) using $\,\textrm{SO}(6,3)\,$ generators in the light-like basis. This matrix is different from the counterpart matrix $\,\mathcal{N}\,$ given in (\ref{GaugeKinMat}) and (\ref{GaugeKinMatBlocks}) which appears upon direct truncation of the $\,\textrm{ISO}(7)\,$ maximal supergravity of \cite{Guarino:2015qaa} to its $\,\textrm{SO}(3)_{\textrm{R}}$-invariant sector.

Following the general discussion of \cite{deWit:2005ub}, two Lagrangians with matrices $\,\mathcal{N}\,$ and $\,\mathcal{N}'\,$ being related by a non-linear transformation
\begin{eqnarray}
\cN' = \big( C + D \cN \big)  \big( A + B \cN \big)^{-1}  \ ,
\end{eqnarray}
defined in terms of a matrix 
\begin{eqnarray}
\label{S_matrix}
{\cal S} =
\left(
\begin{array}{cc}
A & B  \\
C & D   \\
\end{array}
\right) \in \textrm{Sp}(18,\mathbb{R}) \ ,
\end{eqnarray}
are equivalent by electromagnetic duality. It is then straightforward to check that, firstly, the matrix $\,\mathcal{S}\,$ in (\ref{S_matrix}) with
\begin{equation} 
A =  - \tfrac{1}{\sqrt{2}}
\left(
\begin{array}{ccc}
0 & 0  & 0 \\
0 & \mathbb{I}_3 & 0  \\
0 & 0 &  0
\end{array}
\right) 
\,,\, 
B = - C = 
\left(
\begin{array}{ccc}
\mathbb{I}_3 & 0  & 0 \\
0 & 0 & 0  \\
0 & 0 &  -\mathbb{I}_3
\end{array}
\right)
\,,\, 
D = - \sqrt{2} 
\left(
\begin{array}{ccc}
0 & 0  & 0 \\
0 & \mathbb{I}_3 & 0  \\
0 & 0 &  0
\end{array}
\right)  ,
\end{equation} 
leaves invariant $\,\Omega_{\textrm{Sp}(18,\mathbb{R})}\,$ in (\ref{Omega_N8_truncation}) and, secondly, it acts non-linearly on the matrix $\,\cN\,$ in (\ref{GaugeKinMat}) and (\ref{GaugeKinMatBlocks}) and brings it to the matrix $\,\cN'\,$ in (\ref{NkinSW}). This change of symplectic frame, combined with our choice of coordinates for $\,\textrm{SO}(6,3)$, are responsible for the  lengthy topological term in (\ref{Ltop_SO(3)_R}) compared to that of (\ref{L_SO(3)_R_N8}). Note also that, while only three magnetic vectors enter the gauge connection in (\ref{CovDers}), six of them do it in the covariant derivatives of (\ref{DM_SO(3)_R}). As a result, a larger number of tensor fields are required in the canonical (and equivalent) $\,\mathcal{N}=4\,$ formulation of the $\,\textrm{SO}(3)_{\textrm{R}}$-invariant sector of the $\,\textrm{ISO}(7)\,$ maximal supergravity.

\section{Further truncations and $\,\cN \ge 2\,$ supersymmetric vacua}
\label{sec:further_trunc}

The $\,\textrm{SO}(3)_{\textrm{R}}$-invariant sector we have just discussed contains, as further subtruncations, the $\,\textrm{SU}(3)\,$ and $\,\textrm{SO}(4)_{\textrm{d} \times \textrm{R}}\,$ invariant sectors of the $\,\textrm{ISO}(7)\,$ maximal supergravity constructed in \cite{Guarino:2015qaa}\footnote{The subgroup $\,\textrm{SO}(4)_{\textrm{d} \times \textrm{R}}\,$ was simply denoted $\,\textrm{SO}(4)\,$ in \cite{Guarino:2015qaa}. Here we attach the label $_{\textrm{d} \times \textrm{R}}$ in order to avoid confusion with the $\,\textrm{SO}(4)\,$ subgroup appearing in (\ref{G-chain}).}. It also contains all known AdS$_{4}$ vacua of the maximal theory (see table~$1$ of \cite{Guarino:2015qaa} for a summary), in particular the $\mathcal{N}=2$ \cite{Guarino:2015jca} and $\mathcal{N}=3$ \cite{Gallerati:2014xra} vacua. The distinctive feature of the $\,\textrm{SO}(3)_{\textrm{R}}$-invariant sector is that supersymmetric AdS$_{4}$ vacua, especially the $\mathcal{N}=2$ and $\mathcal{N}=3$ ones, exhibit \textit{all} their supersymmetries within this half-maximal sector of the $\,\textrm{ISO}(7)\,$ maximal supergravity. This will be verified by locating their scalar and vector mass spectra within this sector into the multiplet structure of the corresponding $\textrm{OSp}(4|\mathcal{N})$ supergroup.

\subsection{$\textrm{SU}(3)$-invariant sector}

The $\,\textrm{SU}(3)$-invariant sector of the $\,\textrm{ISO}(7)\,$ maximal supergravity describes an $\,\cN=2\,$ gauged supergravity coupled to a vector multiplet and the universal hypermultiplet, with an abelian $\,\textrm{G}=\mathbb{R} \times \textrm{U}(1)\,$ gauging of the universal hypermultiplet \cite{Guarino:2015qaa}. The $\,2 \, + \, 4\,$ scalars in this sector were denoted $\,( \varphi \, , \, \chi )\,$ and $\,( \phi \, , \, a \, , \, \zeta \, , \, \tilde \zeta)$ in  \cite{Guarino:2015qaa}, and are recovered from the $\,\textrm{SO}(3)_{\textrm{R}}$-invariant sector upon identifying $\,( \varphi \, , \, \chi )\,$ with the $\,\textrm{SL}(2)\,$ axion-dilaton in (\ref{M_alphabeta}) and furthermore
\begin{eqnarray}
\label{N=4toSU3inv}
{\bm \nu} = 
\left(
\begin{array}{ccc}
e^{- \phi} & 0  & 0 \\[5pt]
0 & e^{- \phi} & 0   \\[5pt]
0 & 0 & e^{- \varphi}
\end{array}
\right)
\,\,\, , \,\,\,
{\bm a} = 
\left(
\begin{array}{ccc}
0 & -a  & 0 \\[5pt]
a & 0 & 0   \\[5pt]
0 & 0 & 0
\end{array}
\right)
\,\,\, , \,\,\,
{\bm b} = - \tfrac{1}{\sqrt{2}}
\left(
\begin{array}{ccc}
\zeta & -\tilde{\zeta}  & 0 \\[5pt]
\tilde{\zeta} & \zeta & 0   \\[5pt]
0 & 0 & 2 \,  \chi
\end{array}
\right)  .
\end{eqnarray} 
The electric and magnetic vector fields were denoted $\,(A^0 \, , \, A^1 \, , \, \tilde{A}_0 \, , \, \tilde{A}_1)\,$ in \cite{Guarino:2015qaa}, and are identified as
\begin{equation} 
\label{N=4toSU3invVecs}
A'^i = \left(\begin{array}{c} 0 \\ 0 \\ A^1 \end{array} \right)
\hspace{3mm} , \hspace{3mm}
A^{(\textrm{L}) a} = \left(\begin{array}{c} 0 \\ 0 \\ -2 \, A^1 \end{array} \right)
\hspace{3mm} , \hspace{3mm}
A^{(\textrm{t}) i} = \left(\begin{array}{c} 0 \\ 0 \\ - A^0 \end{array} \right) \ ,
\end{equation}
together with
\begin{equation} 
\label{N=4toSU3invVecsMagnetic}
\tilde{A}'_{i} = \left(\begin{array}{c} 0 \\ 0 \\ \tfrac{1}{3} \,  \tilde{A}_1 \end{array} \right)
\hspace{3mm} , \hspace{3mm}
\tilde{A}^{(\textrm{L})}_{a} = \left(\begin{array}{c} 0 \\ 0 \\ -\tfrac{1}{3} \, \tilde{A}_1 \end{array} \right)
\hspace{3mm} , \hspace{3mm}
\tilde{A}^{(\textrm{t})}_{i} = \left(\begin{array}{c} 0 \\ 0 \\ - \tilde{A}_0 \end{array} \right) \ .
\end{equation}
With these identifications, the $\,\cN=4\,$ supergravity here reduces to the $\,\mathcal{N}=2\,$ model presented in sec.~$3$ of \cite{Guarino:2015qaa}. Therefore, all the AdS$_{4}$ vacua in the $\,\textrm{SU}(3)$-invariant sector (see Table~$3$ in \cite{Guarino:2015qaa}) are also vacua in the $\,\textrm{SO}(3)_{\textrm{R}}$-invariant sector.

A particularly interesting solution is an AdS$_{4}$ vacuum preserving $\,\cN=2\,$ supersymmetry and $\,\textrm{G}_{0}=\textrm{U}(1)\,$ residual gauge symmetry within the $\,\textrm{SU}(3)$-invariant sector. It has non-vanishing scalars
\begin{equation}
\label{N=2vevs}
e^{6\varphi} =  \tfrac{64}{27} \left( \tfrac{g}{m} \right)^2
\hspace{5mm} , \hspace{5mm}
\chi = -\tfrac{1}{2} \left(\tfrac{m}{g} \right)^\frac13
\hspace{5mm} , \hspace{5mm}
e^{6\, \phi} =   8 \left( \tfrac{g}{m} \right)^2
\hspace{5mm} , \hspace{5mm}
a = \zeta = \tilde{\zeta} = 0  \ .
\end{equation} 
An explicit computation of the $\,\mathcal{N}=4\,$ gravitino mass terms following \cite{Schon:2006kz} shows that this vacuum also preserves $\,\cN=2\,$ supersymmetry within the $\,\textrm{SO}(3)_{\textrm{R}}$-invariant sector, as an analysis of the spectrum confirms. The scalar squared masses $M^2L^2$, normalised to the AdS$_{4}$ radius $L$, read
\begin{equation} 
\label{N=2spectrum_N2}
\big( 3 +\sqrt{17} \, , \,  2 \, , \,  2 \, , \,  2 \, , \,  3 -\sqrt{17} \big) \; , \; \big( - 2 \, , \, - 2  \big) \; , \; 2 \times \big( - \tfrac{20}{9} \, , \, - \tfrac{14}{9}  \big) \; , \; 2 \times \big( - \tfrac{14}{9}  \big)  \; , \; 7 \times 0 \ .
\end{equation}
These can be allocated into the following OSp$(4|2)$ multiplets, from left to right: one long massive vector multiplet, one massless vector multiplet, two hypermultiplets and two short gravitino multiplets. The zero eigenvalues correspond to the Goldstone bosons eaten up by vectors that become massive after the symmetry breaking $\,\textrm{G}=\textrm{ISO}(3)' \times \textrm{SO}(3)_{\textrm{L}} \rightarrow {\textrm{G}_{0}=\textrm{U}(1)' \times \textrm{U}(1)_{\textrm{L}}}\,$. This is confirmed by the vector squared masses, $M^2L^2$,
\begin{equation}
\label{N=2spectrum_N2_vec}
1 \, \times \,  4 
\,\,\,\, , \,\,\,\, 
2 \, \times \,  \tfrac{28}{9} 
\,\,\,\, , \,\,\,\, 
4 \, \times \, \tfrac{4}{9}
\,\,\,\, , \,\,\,\, 
2 \, \times \,  0  \ .
\end{equation}
The OSp$(4|2)$ multiplet structure of the full $\,\cN=8\,$ dyonically-gauged $\,\textrm{ISO}(7)\,$ supergravity at this vacuum can be found in tables $3$ and $4$ of \cite{Pang:2017omp}.

\subsection{$\textrm{SO}(4)_{\textrm{d} \times \textrm{R}}$-invariant sector}

The $\,\textrm{SO}(4)_{\textrm{d} \times \textrm{R}} \sim \textrm{SO}(3)_{\textrm{d}} \times \textrm{SO}(3)_{\textrm{R}}\,$ invariant sector of the $\,\textrm{ISO}(7)\,$ maximal supergravity, with $\,\textrm{SO}(3)_{\textrm{d}}\,$ being the diagonal subgroup in $\,\textrm{SO}(3)' \times \textrm{SO}(3)_{\textrm{L}}\,$, describes $\,\cN=1\,$ supergravity coupled to two chiral multiplets and no vector multiplets~\cite{Guarino:2015qaa}. We denote the four scalars in this sector $\,( \phi'  \, , \, \rho')\,$ and $\,( \varphi' \, , \, \chi' )\,$ (no primes were used in \cite{Guarino:2015qaa}) which are recovered from the $\,\textrm{SO}(3)_{\textrm{R}}$-invariant sector upon identification of the $\,\textrm{SL}(2)\,$ axion-dilaton in (\ref{M_alphabeta}) as $\,(\varphi \, , \, \chi) = (\phi' \, , \, \rho')\,$ and
\begin{eqnarray}
\label{N=4toSO4inv}
{\bm \nu} = 
\left(
\begin{array}{ccc}
e^{- \varphi'} & 0  & 0 \\[5pt]
0 & e^{- \varphi'} & 0   \\[5pt]
0 & 0 & e^{- \varphi'}
\end{array}
\right)
\,\,\,\,\,\,\, , \,\,\,\,\,\,\,
{\bm a} = 0
\,\,\,\,\,\,\, , \,\,\,\,\,\,\,
{\bm b} = - \sqrt{2}
\left(
\begin{array}{ccc}
\chi' & 0 &  0 \\[5pt]
0 & \chi' & 0   \\[5pt]
0 & 0 & \chi'
\end{array}
\right) \ .
\end{eqnarray} 
With these identifications, and setting to zero all the vector and tensor fields, the $\,\cN=4\,$ supergravity here reduces to the $\,\cN=1\,$ model presented in sec.~$5$ of \cite{Guarino:2015qaa}. As a result, all the AdS$_{4}$ vacua in the $\,\textrm{SO}(4)_{\textrm{d} \times \textrm{R}}$-invariant sector (see Table~$4$ in \cite{Guarino:2015qaa}) are again vacua in the $\,\textrm{SO}(3)_{\textrm{R}}$-invariant sector.

Within this $\,\cN=1\,$ model there is a non-supersymmetric but perturbatively stable AdS$_{4}$ vacuum located at
\begin{equation}
\label{N=3vevs}
e^{6 \phi'} =  \tfrac{4}{27} \left( \tfrac{g}{m} \right)^2 
\hspace{3mm} , \hspace{3mm}
\rho' = -2^{-\frac13} \left(\tfrac{m}{g} \right)^\frac13
\hspace{3mm} , \hspace{3mm}
e^{6 \varphi'} = \tfrac{256}{27}  \left( \tfrac{g}{m} \right)^2
\hspace{3mm} , \hspace{3mm}
\chi' =  2^{-\frac43} \left(\tfrac{m}{g} \right)^\frac13  \ . 
\end{equation} 
The importance of this vacuum relies on the fact that it preserves $\,\cN=3\,$ supersymmetry within the $\,\textrm{SO}(3)_{\textrm{R}}$-invariant sector. We have verified this by direct computation of the $\,\mathcal{N}=4\,$ gravitino mass terms following \cite{Schon:2006kz}, and by analysing the scalar mass spectrum (normalised to the AdS$_{4}$ radius)
\begin{eqnarray} 
\label{N=2spectrum_N3}
3 \, (1+ \sqrt{3}) \, , \quad   6 \times \big( 1+ \sqrt{3} \big) \, , \quad   6 \times \big( 1-  \sqrt{3} \big) \, , \quad  3 \, (1- \sqrt{3}) \, , \quad    6 \times 0 \  .
\end{eqnarray}
The non-zero masses fill out a long gravitino multiplet of OSp$(4|3)$ (see table $2$ of \cite{Fre:1999gok} with $J_0= 0$, $E_0 = \sqrt{3}$). The six zero masses correspond to the Goldstone bosons that are eaten up by the vectors that become massive after the symmetry breaking $\,{\textrm{G}=\textrm{ISO}(3)' \times \textrm{SO}(3)_{\textrm{L}} \rightarrow \textrm{G}_{0}=\textrm{SO}(3)_{\textrm{d}}}\,$.  This is again confirmed by the vector masses (normalised to the AdS$_{4}$ radius) 
\begin{equation}
\label{N=2spectrum_N3_vec}
3 \, \times \, \big( 3 \pm \sqrt{3} \big) 
\,\,\,\, , \,\,\,\, 
3 \, \times \,  0 \ .
\end{equation}

\section{New non-supersymmetric vacua}
\label{sec:vacua}

In ref.~\cite{Guarino:2015qaa}, the vacua of dyonic $\,\textrm{ISO}(7)$-gauged maximal supergravity with at least $\,\textrm{SU}(3)\,$ or $\,\textrm{SO}(4)_{\textrm{d} \times \textrm{R}}\,$ invariance were investigated, giving rise to four different supersymmetric vacua and six non-supersymmetric ones. Of the latter, two of them are perturbatively unstable in the sense of displaying normalised scalar masses below the BF bound for stability in $\,\textrm{AdS}_4\,$, and the remaining four are stable. We can extend this classification now by considering the $\,\textrm{SO}(3)_\textrm{R}\,$-invariant potential of (\ref{V_SO3}). 

To do this we have performed a numeric procedure involving the minimisation of the scalar potential, in field space, by a Newton method. To generate the initial values for the 17-dimensional parameter space (recall that there are three St\"uckelberg scalars $\,a_{ij}\,$ that do not enter (\ref{V_SO3})) we considered a uniform distribution consisting of a ``hypercube" in 17-dimensional space with $\,|\varphi_i|\leq1\,$ for each of the 17 fields $\,\varphi_i\,$, with approximately 50000 initial points being generated. To extend the search we considered ``hypercubes" of increasing size, and noticed that the Newton method started to be inefficient when $\,|\varphi_i|\approx 2\,$. We produced approximately 20000 initial points in these larger parameter spaces.

The minimisation process gave rise to critical values of the fields, $\,\varphi_i^*\,$, that we used to calculate the normalised mass spectrum of scalar fluctuations around the said minimum and the $\,\mathcal{N}=4\,$ gravitino (squared) mass matrix. The eigenvalues of the latter, $\,m^2_a\,$, determine the number of supersymmetries preserved by the vacuum by counting the number of eigenvalues satisfying $\,m^2_a = -3V_*/4\,$, with $\,V_*\,$ the value of the potential evaluated at $\,\varphi_i^*\,$.

With this method we reproduce all the fixed points listed in \cite{Guarino:2015qaa} and find five new non-supersymmetric vacua with values of the scalar potential (setting $\,g=c=1\,$)
\begin{equation}
 V_* = \{ -21.867393 \ ,  \quad -23.322349 \ , \quad  -23.456053 \ , \quad  -23.456098 \ , \quad  -23.458780   \} \ .
\end{equation}
While the vacuum corresponding to the largest value of the potential is perturbatively unstable, the remaining four are stable within the $\,\textrm{SO}(3)_{\textrm{R}}$-invariant sector. For these vacua, it would be interesting to perform an analysis of perturbative stability within the full ISO(7) maximal theory. This is beyond the scope of this work.

We proceed now to list the position in scalar field space, as well as the vector and scalar normalised mass spectra, of these vacua.

\subsubsection*{\normalfont{$i)$ \textit{Vacuum with $\,V_* = -21.867393\,$}}}

The position of this vacuum in the scalar manifold is given by
\begin{align}
\chi & = -0.715397, \quad \varphi = 0.0795045, \quad \phi_i = \left(
\begin{array}{c}
 0.640745 \\
 0.575310 \\
 0.0475177 \\
\end{array}
\right) , \\ 
h^i{_j} & = \left(
\begin{array}{ccc}
 * & -0.0150532 & 0.193075 \\
 * & * & -0.521471 \\
 * & * & * \\
\end{array}
\right), \,\,\,\, (i <j) \ ,\\[2mm]
b^a{_j} & = \left(
\begin{array}{ccc}
 -0.451841 & -0.226962 & -0.0251816 \\
 0.0839092 & 0.0421481 & 0.00467636 \\
 -0.269282 & -0.135262 & -0.0150074 \\
\end{array}
\right) \ .
\end{align}
The normalised vector masses read
\begin{align}
M^2 L^2 & = \{5.61795, \ 3.84978, \ 3.66158, \ 0.744392, \ 0.663310, \nonumber \\
&\qquad \ 0.521957 \, (\times 2), \ 0.274647, \ 0    \} \ ,
\end{align}
so there is a massless vector and the gauge group is broken to an abelian $\,\textrm{U}(1)\,$. The normalised scalar masses are given by
\begin{align}
M^2 L^2 & = \{-2.66511, \ -2.63847, \ -2.09157, \ -1.54651, \ -1.08210 \, (\times 2), \nonumber \\
&\qquad 0 \, (\times 8), \quad 0.732955, \ 2.63847, \ 4.17512, \ 4.18428 \,  (\times 2), \ 7.01842\} \ ,
\end{align}
so that the vacuum is perturbatively unstable. Note the presence of eight massless scalars being Goldstone bosons eaten up by the massive vectors.

\subsubsection*{\normalfont{$ii)$ \textit{Vacuum with $\,V_* = -23.322349\,$}}}

The position of this vacuum in the scalar manifold is given by
\begin{align}
\chi & = -0.861587, \quad\varphi = -0.300962, \quad \phi_i = \left(
\begin{array}{c}
 0.505757 \\
 0.497103 \\
 0.556115 \\
\end{array}
\right), \\ 
h^i{_j} & = \left(
\begin{array}{ccc}
 * & -0.184685 & 0.136010 \\
 * & * & -0.139564 \\
 * & * & * \\
\end{array}
\right), \,\,\,\, (i <j)  \ ,\\[2mm]
b^a{_j} & = \left(
\begin{array}{ccc}
 0.457458 & 0.304725 & -0.160877 \\
 -0.127647 & -0.0924051 & -0.526954 \\
 0.278269 & -0.440951 & -0.0597982 \\
\end{array}
\right) \ .
\end{align}
The normalised vector masses read
\begin{align}
M^2 L^2 & = \{5.61055, \ 4.51560 \, (\times 2), \ 1.26025, \ 1.23284 \, (\times 2), \ 0.0591156 \, (\times 2), \ 0    \} \ ,
\end{align}
so there is a massless vector and the gauge group is again broken to $\,\textrm{U}(1)\,$. The normalised scalar masses are given by
\begin{align}
M^2 L^2 & = \{-2.20094, \ -1.6884 \, (\times 2), \ -0.868452, \ 0 \, (\times 8), \ 0.885771,  \nonumber \\
& \qquad  1.98171 \, (\times 2), \ 2.26671, \ 3.60935 \, (\times 2), \ 4.41937, \ 8.16298\} \ ,
\end{align}
so that the vacuum is this time perturbatively stable. Note again the presence of eight massless scalars being Goldstone bosons eaten up by the massive vectors.

\subsubsection*{\normalfont{$iii)$ \textit{Vacuum with $\,V_* = -23.456053\,$}}}

The position of this vacuum in the scalar manifold is given by
\begin{align}
\chi & = 0.273189, \quad \varphi = 0.0399038, \quad \phi_i = \left(
\begin{array}{c}
 0.461474 \\
 0.659100 \\
 0.653142 \\
\end{array}
\right), \\ 
h^i{_j} & = \left(
\begin{array}{ccc}
 * & 0.0534187 & -0.0707990 \\
 * & * & -0.0149067 \\
 * & * & * \\
\end{array}
\right), \,\,\,\, (i <j)  \ , \\[2mm]
b^a{_j} & = \left(
\begin{array}{ccc}
 -0.516908 & -0.154388 & -0.249397 \\
 -0.0485321 & -0.470547 & 0.209088 \\
 0.462023 & -0.114350 & -0.400746 \\
\end{array}
\right) \ .
\end{align}
The normalised vector masses read
\begin{align}
M^2 L^2 & = \{4.29451, \ 3.62600 \, (\times 2), \ 2.66757 \, (\times 2), \ 2.29373, \ 0.0883439 \, (\times 2), \ 0    \} \ ,
\end{align}
so there is a massless vector and the gauge group is also broken to $\,\textrm{U}(1)\,$. The normalised scalar masses are given by
\begin{align}
M^2 L^2 & = \{-1.58248, \ -1.36606 \, (\times 2), \ -0.987543, \ -0.920226 \, (\times 2), \ -0.145118, \nonumber \\
& \qquad -0.0277198 \, (\times 2), \ 0 \, (\times 8), \ 1.14647, \ 5.78023, \ 6.29251\} \ ,
\end{align}
so that the vacuum is perturbatively stable. Note also the presence of eight massless scalars being Goldstone bosons eaten up by the massive vectors.

\subsubsection*{\normalfont{$iv)$ \textit{Vacuum with $\,V_* = -23.456098\,$}}}

The position of this vacuum in the scalar manifold is given by
\begin{align}
\chi & = 0.267119, \quad \varphi = 0.0383881, \quad \phi_i = \left(
\begin{array}{c}
 0.500397 \\
 0.649735 \\
 0.623964 \\
\end{array}
\right), \\ 
h^i{_j} & = \left(
\begin{array}{ccc}
 * & 0.0622572 & -0.108150 \\
 * & * & -0.0330283 \\
 * & * & * \\
\end{array}
\right), \,\,\,\, (i <j)  \ , \\[2mm]
b^a{_j} & = \left(
\begin{array}{ccc}
 -0.238203 & -0.0134332 & 0.553237 \\
 -0.584557 & -0.259510 & -0.0347291 \\
 -0.197645 & 0.446453 & -0.0379982 \\
\end{array}
\right) \ .
\end{align}
The normalised vector masses read
\begin{align}
M^2 L^2 & = \{4.26513, \ 3.70813, \ 3.59326, \ 2.67942, \ 2.63558, \nonumber \\ 
& \qquad \ 2.30724, \ 0.0948198,  \ 0.0753444, \ 0.00171607  \} \ ,
\end{align}
so all vectors become massive and the gauge group is fully broken at this vacuum. The normalised scalar masses are given by
\begin{align}
M^2 L^2 & = \{ -1.58451, \ -1.35168, \ -1.34552, \ -0.977451, \ -0.96062, \ -0.874352,  \nonumber \\
& \qquad   -0.214610, \ 0 \, (\times 9) ,  \ 0.0395421,  \, 1.14238,    \ 5.77263,  \ 6.29845   \} 
\end{align}
so that the vacuum is perturbatively stable. Note this time the presence of nine massless scalars being Goldstone bosons eaten up by the massive vectors.

\subsubsection*{\normalfont{$v)$ \textit{Vacuum with $\,V_* = -23.458780\,$}}}

The position of this vacuum in the scalar manifold is given by
\begin{align}
\chi & = 0.373634, \quad \varphi = 0.0963292, \quad \phi_i = \left(
\begin{array}{c}
 0.560889 \\
 0.489926 \\
 0.692913 \\
\end{array}
\right), \\ 
h^i{_j} & = \left(
\begin{array}{ccc}
 * & 0.261188 & -0.0655070 \\
 * & * & -0.0769042 \\
 * & * & * \\
\end{array}
\right), \,\,\,\, (i <j)  \ , \\[2mm]
b^a{_j} & = \left(
\begin{array}{ccc}
 -0.331197 & -0.378733 & -0.264801 \\
 0.427584 & 0.105189 & -0.346559 \\
 -0.126002 & 0.476769 & -0.226711 \\
\end{array}
\right) \ .
\end{align}
The normalised vector masses read
\begin{align}
M^2 L^2 & = \{4.59692, \ 3.26745 \, (\times 2), \ 2.82061 \, (\times 2), \ 2.16737, \ 0.132118 \, (\times 2), \ 0    \} \ ,
\end{align}
so there is a massless vector and the gauge group is again broken to $\,\textrm{U}(1)\,$. The normalised scalar masses are given by
\begin{align}
M^2 L^2 & = \{-1.59719 \, (\times 2), \ -1.58649, \ -1.22285, \ -0.935656 \, (\times 2),  \nonumber \\ 
& \qquad -0.480612 \, (\times 2), \ 0 \, (\times 8), \ 0.0852111, \ 1.16111, \ 5.8955, \ 6.22349\} \ ,
\end{align}
so that the vacuum is also perturbatively stable. Note the presence of eight massless scalars being Goldstone bosons eaten up by the massive vectors.

\section{Discussion} \label{sec:Discussion}

We have truncated $\,D=4\,$ and $\,\cN=8\,$ dyonic ISO(7) supergravity \cite{Guarino:2015qaa} to its $\,\textrm{SO}(3)_{\textrm{R}}$-invariant sector. This corresponds to $\,D=4\,$ and $\,\cN=4\,$ supergravity coupled to three vector multiplets with an $\,\textrm{ISO}(3)' \times \textrm{SO}(3)_{\textrm{L}}\,$ gauge group. We have also cast this model in canonical $\,\cN=4\,$ form \cite{Schon:2006kz}, and have shown that the resulting gauging is new from the $\,\cN=4\,$ perspective in that it contains matter that is charged under both factors of the gauge group. Since $\,\textrm{SO}(3)_{\textrm{R}}\,$ is contained both in the $\,\textrm{SU}(3)\,$ and the $\,\textrm{SO}(3)_{\textrm{d}} \times \textrm{SO}(3)_{\textrm{R}}\,$ subgroups of $\,\textrm{SO}(7) \subset \textrm{ISO}(7)\,$, our model encompasses those sectors, constructed in \cite{Guarino:2015qaa}, that are invariant under the latter subgroups. Thus, the $\,\cN=4\,$ and $\,\textrm{SO}(3)_{\textrm{R}}$-invariant sector contains all known supersymmetric and non-supersymmetric critical points of the $\,\textrm{ISO}(7)\,$ maximal supergravity and, as we show in section~\ref{sec:vacua}, five numerical non-supersymmetric points that are new. In particular, this sector contains the two points of $\,\textrm{ISO}(7)\,$ supergravity with the largest possible supersymmetry: the $\,\cN=2\,$ and $\,\cN=3\,$ vacua of \cite{Guarino:2015jca} and \cite{Gallerati:2014xra}. These arise as supersymmetric vacua and exhibit their full $\,\cN=2\,$ and $\,\cN=3\,$ supersymmetries already within our $\,\cN=4\,$ model.

The $\,\cN=2\,$ and $\,\cN=3\,$ vacua are dual to superconformal Chern-Simons theories with the same supersymmetries, that were respectively described in \cite{Guarino:2015jca} and \cite{Gaiotto:2007qi}. These have adjoint matter in the $\,\bm{3}\,$ and the $\,\bm{2}\,$ of the corresponding flavour groups, $\,\textrm{SU}(3)\,$ and $\,\textrm{SO}(3)_{\textrm{R}}\,$. It was argued in the latter reference that adding a mass term for one of the $\,\bm{3}\,$ adjoint chirals should cause the former theory to flow into the latter. The low energy flavour group $\,\textrm{SO}(3)_{\textrm{R}}\,$ must be preserved along the flow. For this reason, if this picture is correct, the $\,\textrm{SO}(3)_{\textrm{R}}$-invariant sector that we have constructed in this paper must not only contain the $\,\cN=2\,$ and $\,\cN=3\,$ endpoints of the flow, but also  a full $\,\textrm{SO}(3)_{\textrm{R}}$-invariant domain-wall solution interpolating between both of them. We have verified that this is indeed the case, in agreement with the argument put forward in appendix~C.3 of \cite{Guarino:2016ynd}, by explicitly constructing the interpolating solution. The details will be reported in \cite{Guarino:2019snw}.

\section*{Acknowledgements}

AG is partially supported by the Spanish government grant MINECO-16-FPA2015-63667-P and by the Principado de Asturias through the grant FC-GRUPIN-IDI/2018/000174.  JT is supported  by the European Research Council, grant no. 725369.  OV is supported by the NSF grant PHY-1720364 and, partially, by grants SEV-2016-0597, FPA2015-65480-P and PGC2018-095976-B-C21 from MCIU/AEI/FEDER, UE. Finally, AG and JT thank the Instituto de F\'isica Te\'orica (IFT) UAM-CSIC for hospitality during the completion of this work.

\appendix

\section{Group-theoretical embedding of $\,\textrm{SO}(3)_{\textrm{R}} \subset \textrm{E}_{7(7)}\,$} 
\label{sec:N8_embedding}

Let us consider the $\,\textrm{SO}(3)_{\textrm{R}}\,$ subgroup characterised by the embedding (lower chain) (\ref{Embedding_SO3_R}) and introduce a set of constant $\,4 \times 4\,$ matrices $\,J^{a}_\pm\,$ with $\,a=1,2,3\,$. The components of these matrices $\,(J^{a}_\pm)_{\alpha \beta}\,$ with $\,\alpha = (0 , a )\,$ are given by\footnote{The index $\,\alpha\,$ in this appendix should not be confused with the $\,\textrm{SL}(2)\,$ index $\,\alpha=\pm\,$ in the main text.}
\begin{equation}
(J^{a}_\pm)_{0 b } = \mp \delta^{a}_{b} 
\hspace{8mm} , \hspace{8mm}
(J^{a}_\pm)_ {bc  }  = -\epsilon^{a}{}_{bc }   \ ,
\end{equation}
so that they are antisymmetric and (anti)-self-dual
\begin{equation}
\label{antiSD}
(J^{a}_\pm)_{\alpha \beta} = \pm \tfrac12 \epsilon_{\alpha \beta \gamma \delta} \, (J^{a}_\pm)^{\gamma \delta} \ ,
\end{equation}
satisfy the quaternion algebra
\begin{equation}
\label{quaternion}
({J^{a}_\pm})^{\alpha}{}_{\gamma} ({J^b}_\pm)^{\gamma}{}_{\beta}  = -\delta^{ab } \delta^\alpha_\beta + \epsilon^{ab }{}_{c} (J^{c}_\pm)^\alpha{}_\beta \ ,
\end{equation}
and the identity
\begin{equation}
\label{QuadID}
(J^{a}_\pm)_{\alpha \beta} (J_{ a \pm} )_{\gamma \delta}  = 2 \delta_{\alpha[\gamma}\delta_{\delta]\beta} \pm  \epsilon_{\alpha \beta \gamma \delta} \ .
\end{equation} 
Indices $\,a\,$ and $\,\alpha\,$ are raised and lowered with $\,\delta_{ab}\,$ and $\,\delta_{\alpha \beta}\,$.

Let us now consider the generators $\,t_A{}^B\,$ and $\,t_{ABCD}\,$ of $\,\textrm{E}_{7(7)}\,$ in the $\,\textrm{SL}(8)\,$ basis\footnote{We follow the conventions in appendix~C of \cite{Guarino:2015qaa}.} and split the fundamental $\,\textrm{SL}(8)\,$ index $\,A=1 , \ldots, 8\,$ as $\,A= (i , \alpha, 8)\,$. The indices $\,i\,$ and $\,\alpha\,$ refer to the fundamental representations of $\,\textrm{SO}(3)'\,$ and $\,\textrm{SO}(4)\,$ in (\ref{Embedding_SO3_R}), respectively, whereas $\,a\,$ denotes the fundamental representation of $\,\textrm{SO}(3)_{\textrm{L}}\,$ and similarly for $\,\textrm{SO}(3)_{\textrm{R}}\,$. Then the $\,\textrm{SO}(3)'\,$,  $\,\textrm{SO}(3)_{\textrm{L}}\,$ and  $\,\textrm{SO}(3)_{\textrm{R}}\,$ subgroups of $\,\textrm{E}_{7(7)}\,$ are respectively generated by
\begin{eqnarray} 
\label{GensSO3s}
G^\prime_i = \epsilon_{ij}{}^k \, t_k{}^j \qquad , \qquad
L_{a} = - \tfrac12 (J_{a - } )_\alpha{}^\beta \,  t_\beta{}^\alpha \qquad , \qquad
R_{a} =  \tfrac12 (J_{a + } )_\alpha{}^\beta \,  t_\beta{}^\alpha \  ,
\end{eqnarray}
and have non-vanishing commutation relations of the form
\begin{equation} 
\label{SO3comm}
[ G^\prime_i \,,\, G^\prime_j ] = \epsilon_{ij}{}^k G^\prime_k \qquad  , \qquad 
[ L_{a} \,,\, L_{b}  ] = \epsilon_{ ab}{}^{c} L_{c} \qquad  , \qquad 
[ R_{a} \,,\, R_{b}  ] = -\epsilon_{ ab }{}^{c} R_{c} \  .
\end{equation}
The $\,\textrm{SO}(3)_\textrm{d}\,$ diagonal subgroup inside $\,\textrm{SO}(3)^\prime \times \textrm{SO}(3)_{\textrm{L}}\,$ is generated by $\,D_i \equiv G^\prime_i + \delta^{b}_i \, L_{b}\,$. 
\\

\noindent\textbf{Scalar sector:}  The group $\,\textrm{SO}(3)_\textrm{R}\,$ commutes with $\,\textrm{SL}(2, \mathbb{R})  \times  \textrm{SO}(6,3)\,$ inside $\,\textrm{E}_{7(7)}\,$. To see this, note that the generators $R_{a}$ in (\ref{GensSO3s}) commute, on the one hand, with
\begin{eqnarray} 
\label{solvSL2Gen}
 H_0 \equiv  t_i{}^i  - t_\alpha{}^\alpha  +t_8{}^8 \quad , \qquad E_0 \equiv \tfrac{1}{3 !} \, \epsilon^{ijk} \,  t_{ijk8} \; , 
\end{eqnarray}
together with the negative root $\,E_0^\sharp \equiv \frac{1}{4 !} \epsilon^{\alpha\beta\gamma\delta} t_{\alpha\beta\gamma\delta}\,$ associated with $\,E_0\,$. They also commute, on the other hand, with
\begin{eqnarray} 
\label{solvSO63Gen}
& H_1 \equiv \tfrac{1}{\sqrt{2}} \left( t_1{}^1-t_2{}^2-t_3{}^3+t_8{}^8 \right) \; , \qquad
 H_2 \equiv \tfrac{1}{\sqrt{2}} \left( -t_1{}^1+t_2{}^2-t_3{}^3+t_8{}^8 \right) \; ,  \nonumber \\[5pt]
 & H_3 \equiv \tfrac{1}{\sqrt{2}} \left( -t_1{}^1-t_2{}^2+t_3{}^3+t_8{}^8 \right)  \; ,   \\[5pt] 
& E_i{}^j \equiv -\delta_{ik} \delta^{jh} \, t_h{}^k \; \textrm{(with $i < j $)} \; , \quad 
V^{ij} \equiv \epsilon^{ij}{}_{k} \, t_8{}^k  \; , \quad
U_{a}{}^j \equiv 3\sqrt{2} \, \delta^{jk} \, (J_{a - } )^{\alpha \beta} \, t_{k\alpha \beta 8} \; , \nonumber
\end{eqnarray}
along with $\,L_{a}\,$ defined in (\ref{GensSO3s}) and the negative roots associated with $\,E_i{}^j\,$, $\,V^{ij}\,$ and $\,U_{a}{}^j\,$,
\begin{equation} 
\label{NegRootsSO63Gen}
E_i{}^{j \, \sharp} \equiv - t_i{}^j \; \textrm{(with $i < j $)} \; , \quad 
V^{ij \, \sharp} \equiv \epsilon^{ijk} \, t_k{}^8  \; , \quad
U_{a}{}^{j \, \sharp }  \equiv  - \tfrac{3}{\sqrt{2}} \, \epsilon^{jkl} \, (J_{ a - } )^{\alpha \beta} \, t_{kl\alpha \beta} \; .
\end{equation}
The generators (\ref{solvSL2Gen}) close under commutation as 
\begin{eqnarray}
[H_0  , E_ 0 ] = 4 \, E_0 \; .
\end{eqnarray}
This is well-known to be the commutation relation of the upper triangular, solvable subalgebra of $\,\textrm{SL}(2)\,$ that exponentiates into the first scalar coset in (\ref{Vcosets}). The generators (\ref{solvSO63Gen}) obey commutation relations (no sum over repeated indices) of the form
\begin{eqnarray} 
\label{solvSO63}
&& [H^k , E_i{}^j ] = b^k_{ij} \, E_i{}^j  \; , \qquad 
[H^k , V^{ij} ] = a^k_{ij} \, V^{ij}  \; , \qquad 
[H^k , U_{a}{}^j ] = c_j^k \, U_{a}{}^j  \; , \nonumber \\
&& [ E_i{}^j , E_k{}^\ell ] = \delta_k^j \, E_i{}^\ell - \delta_i^\ell \, E_k{}^j \; , \nonumber \\
&& [ E_i{}^j , V^{k \ell } ] = -\delta_i^k \, V^{j \ell} -  \delta_i^\ell  \, V^{ kj}  \; , \qquad
 [ E_i{}^j , U_{a}{}^k ] = -\delta_i^k \, U_{a}{}^j \; , \nonumber \\ 
&& [U_{a}{}^i ,  U_{b}{}^j ]= \delta_{ a b } V^{ij} \; , 
\end{eqnarray}
where we have defined 
\begin{equation}
a_{ij}^k \equiv \sqrt{2} ( \delta_i^k +   \delta_j^k  ) \; , \qquad
b_{ij}^k \equiv \sqrt{2} ( -\delta_i^k +   \delta_j^k  ) \; , \qquad
c_{i}^k \equiv \sqrt{2} \,   \delta_i^k  \; .
\end{equation}
Commutation relations of the type (\ref{solvSO63}) were studied in \cite{Lu:1998xt} and shown to correspond to the solvable Lie algebras that exponentitate into the coset spaces $\,\textrm{SO}(p,q)/(\textrm{SO}(p) \times \textrm{SO}(q))$. In the present case $\,(p,q)=(6,3)\,$ and the commutation relations in (\ref{solvSO63}) match those in (3.26) of \cite{Lu:1998xt} (with $\,D=7\,$ there and minor notational changes). As a result the generators in (\ref{solvSO63}) exponentiate into the second scalar coset in (\ref{Vcosets}). 

The full coset representative on
\begin{equation}
\mathcal{M}_{\textrm{scalar}} = \frac{\textrm{SL}(2)}{\textrm{SO}(2)} \times \frac{\textrm{SO}(6,3)}{\textrm{SO}(6) \times \textrm{SO}(3)} \ ,
\end{equation}
can therefore be built as
\begin{eqnarray}
{\cal V} &=& e^{12 \chi E_0} \, e^{-\frac14  \varphi H_0}  \, e^{ - b^{a}{}_j U_{a}{}^j} \, e^{ -\frac12  a_{ij} V^{ij} }  \, e^{ - h^i{}_j E_i{}^j} \, e^{ -\frac12 \phi_i H^i } \; , 
\end{eqnarray} 
where the sum on $\,h^i{}_j E_i{}^j\,$ extends only for $\,i<j\,$. The corresponding scalar matrix in maximal supergravity is constructed as $\,{\cal M} = {\cal V} \, {\cal V}^\textrm{T}\,$. The scalar kinetic terms in (\ref{ScalKinTermsSL2}) and (\ref{ScalKinTerms63}), as well as the scalar potential (\ref{V_SO3}), result from substituting $\,{\cal M}\,$ into the general expressions for the parent ISO(7) theory given in sec.~$2$ of \cite{Guarino:2015qaa}. These in turn follow from the general $\,\cN=8\,$ gauged supergravity expressions of \cite{deWit:2007mt}. The kinetic terms in (\ref{ScalKinTerms63}) for the scalars in the vector multiplets match (3.20) of \cite{Lu:1998xt}. 
\\

\noindent\textbf{Vector sector:} The subgroup $\,\textrm{SO}(3)_{\textrm{R}}$ commutes with $\,\textrm{ISO}(3)' \times \textrm{SO}(3)_{\textrm{L}}\,$ inside the $\,\textrm{ISO}(7)\,$ gauge group of the parent $\,\cN=8\,$ theory, namely, $\,\textrm{ISO}(7) \supset \textrm{ISO}(3)' \times \textrm{SO}(3)_{\textrm{L}} \times \textrm{SO}(3)_{\textrm{R}}\,$. This results into
\begin{equation}
\label{G_group_SO(3)_R}
\textrm{G} = \textrm{ISO}(3)' \times \textrm{SO}(3)_{\textrm{L}} =  (\textrm{SO}(3)' \ltimes \mathbb{R}^{3}) \times \textrm{SO}(3)_{\textrm{L}} \ ,
\end{equation}
being the gauge group of the $\,\cN=4\,$ and $\,\textrm{SO}(3)_\textrm{R}$-invariant truncation. As a subgroup, $\,\textrm{G} \subset \textrm{SO}(6,3)\,$ is generated by $\,G^\prime_1 \equiv -(E_2{}^3 - E_2{}^{ 3 \, \sharp}\,$), etc.,  $\,T^i \equiv -\tfrac12 \epsilon^i{}_{jk} V^{jk}\,$ and $\,L_{a}\,$, with the latter and  $G^\prime_i$ defined in (\ref{GensSO3s}). The semidirect action of $\,\textrm{SO}(3)'\,$ on $\,\mathbb{R}^3\,$ is explicitly defined through the commutation relations
\begin{equation}
[ G^\prime_i \, , \, T_j ] = \epsilon_{ij}{}^k T_k \ .
\end{equation}

Let us now determine which of the $\,\cN=8\,$ vector fields gauge the group $\,\textrm{G} \subset \textrm{ISO}(7)$. The gauge fields of the $\,\cN=8\,$ supergravity in \cite{Guarino:2015qaa} were denoted $\,(\cA^{IJ} \, , \, \tilde{\cA}_{IJ})\,$ and $\,(\cA^{I} \, , \, \tilde{\cA}_{I})\,$ with $\,I = 1, \ldots , 7\,$. The former gauge the $\,\textrm{SO}(7)\,$ factor electrically and the latter gauge the $\,\mathbb{R}^7\,$ translations dyonically. In order to identify the subset (\ref{A_fields_SO(3)_R}) of $\,\cN=8\,$ gauge fields that are invariant under $\,\textrm{SO}(3)_\textrm{R}\,$, we split the index $\,I = (i, \alpha)\,$ so that $\,i\,$ and $\,\alpha\,$ respectively label the fundamental of $\,\textrm{SO}(3)'\,$ and $\,\textrm{SO}(4)\,$ in (\ref{Embedding_SO3_R}), and $\,a\,$ labels the fundamental of $\,\textrm{SO}(3)_{\textrm{L}} \subset \textrm{SO}(4)\,$. Then one finds the following identifications for the electric vectors
\begin{equation} 
\label{ElectricVectorsinN=8}
\cA^{ij} = \epsilon^{ij}{}_k \, A^{\prime k} \ , \quad
\cA^{i\alpha} = 0 \ , \quad
\cA^{\alpha \beta} = -\tfrac12  (J_{a - })^{\alpha \beta} \, A^{ (\textrm{L}) a}  \ , \quad
\cA^{i} =  A^{ (\textrm{t}) i} \ ,  \quad
\cA^{\alpha} = 0 \; ,
\end{equation}
and for the magnetic duals
\begin{equation} 
\label{MagneticVectorsinN=8}
\tilde{\cA}_{ij} = \epsilon_{ij}{}^k \, \tilde{A}^{\prime}_{ k} \ , \quad
\tilde{\cA}_{i \alpha} = 0 \ , \quad
\tilde{\cA}_{\alpha \beta} = - (J^{a}{}_{ - })_{\alpha \beta} \, \tilde{A}^{ (\textrm{L}) }_{a}  \ , \quad
\tilde{\cA}_{i} =  \tilde{A}^{ (\textrm{t})}_{ i} \ ,  \quad
\tilde{\cA}_{\alpha} = 0 \; .
\end{equation}
The vectors $\,A^{\prime i}\,$, $\,A^{ (\textrm{t}) i}\,$ and $\,A^{ (\textrm{L}) a}\,$ are gauge fields for, respectively, each of the factor groups in the right hand side of (\ref{G_group_SO(3)_R}), as the field strengths (\ref{N=4electricFS}) confirm.
\\

\noindent\textbf{Two-form sector:} The restricted tensor hierachy of the ISO(7) maximal supergravity includes two-form potentials $\,{\cal B}_I{}^J \equiv \bm{21} + \bm{27}\,$ and $\,{\cal B}^I \equiv \bm{7}^\prime\,$ \cite{Guarino:2015qaa}. The tensor fields $\,{\cal B}_I{}^J\,$ split as 
\begin{eqnarray}
{\cal B}_i{}^j = B_i{}^j \; , \qquad 
{\cal B}_i{}^\alpha = 0 \; , \qquad 
{\cal B}_\alpha{}^i = 0 \; , \qquad 
{\cal B}_\alpha{}^\beta =  B_a \, (J^a_-)_\alpha{}^\beta \; , \qquad 
\end{eqnarray}
whereas the splitting of $\,{\cal B}^I\,$ reads
\begin{eqnarray}
{\cal B}^i = B^i \; , \qquad 
{\cal B}^\alpha = 0 \; . \qquad 
\end{eqnarray}
Importantly, when the magnetic component of the embedding tensor is dualised into a three-form potential (see (\ref{three-forms_magnetic}) below), consistency of the Bianchi identities requires an additional $\,\textrm{SO}(7)$-singlet two-form $\,\mathcal{B}\equiv \textbf{1}\,$ that renders $\,\mathcal{B}_{I}{}^{J}\,$ traceful \cite{Guarino:2015qaa}. Being an $\,\textrm{SO}(7)$-singlet, this additional two-form survives the truncation to the $\,\textrm{SO}(3)_{\textrm{R}}$-invariant sector
\begin{equation}
\label{two-forms_trace}
\mathcal{B} = B \ .
\end{equation}

\noindent\textbf{Three-form sector:} The restricted tensor hierarchy of the $\,\textrm{ISO}(7)\,$ maximal supergravity includes three-form potentials $\,{\cal C}^{IJ} \equiv \bm{1} + \bm{27}\,$ \cite{Guarino:2015qaa}. These have a splitting of the form
\begin{eqnarray}
{\cal C}^{ij} = C^{ij}  \; , \qquad 
{\cal C}^{i \alpha} = 0  \; , \qquad 
{\cal C}^{\alpha \beta} = C^0  \, \delta^{\alpha \beta} \ .
\end{eqnarray}
In addition, there is a three-form potential $\,\tilde{\mathcal{C}} \equiv \bm{1}\,$ dual to the magnetic components of the embedding tensor in the $\,\textrm{ISO}(7)\,$ maximal theory \cite{Guarino:2015qaa} which thus survives the truncation to the $\,\textrm{SO}(3)_{\textrm{R}}$-invariant sector
\begin{equation}
\label{three-forms_magnetic}
\tilde{\mathcal{C}}  = \tilde{C} \ .
\end{equation}
The field strengths associated with these three-form potentials can be used to reconstruct the scalar potential of the truncated theory (see (\ref{VfromH4}) of appendix~\ref{sec:TensHier}).

\section{Tensor hierarchy in the $\,\textrm{SO}(3)_{\textrm{R}}$-invariant sector} 
\label{sec:TensHier}

In the presence of magnetic charges induced by the gauging parameter $\,m\,$, the field strengths of the electric vectors are subject to Bianchi identities
\begin{equation} 
\label{ElectricBianchis_app}
D H^{\prime i}_\2 = 0 \; , \qquad 
D H^{\textrm{(L)} a}_\2 = 0 \; , \qquad 
D H^{\textrm{(t)} i }_\2 = m \, H^{ i }_\3 \; , 
\end{equation}
where $\,H^{ i }_\3\,$ is the three-form field strength of the two-form potential $\,B^i\,$ in (\ref{two-form_fields}). The covariant derivatives in (\ref{ElectricBianchis_app}) are defined as
\begin{equation}
\label{DH_elec_app}
\begin{array}{lll}
D H^{\prime i}_\2 & \equiv & d H^{\prime i}_\2  + g \, \epsilon^i{}_{jk} \, A^{\prime  j } \wedge H^{\prime k}_\2  \ , \\[2mm]
D H_\2^{(\textrm{L}) a } & \equiv & d H_\2^{(\textrm{L}) a } + g \,\epsilon^{a}{}_{ b   c } \, A^{ (\textrm{L})   b  } \wedge H_\2^{(\textrm{L})  c } \ , \\[2mm]
D H_\2^{( \textrm{t})  i } & \equiv & d  H_\2^{( \textrm{t})  i }  +  g \,\epsilon^i{}_{jk} \, A^{\prime  j } \wedge H_\2^{( \textrm{t})  k }   + \epsilon_{ijk} \big( g A^{(\textrm{t}) j } - m \, \delta^{jh} \tilde{A}^{(\textrm{t}) }_h  \big) \wedge  H_\2^{( \textrm{t})  k }  \ .
\end{array}
\end{equation} 
Similarly, even though they do not carry independent dynamics, we can introduce field strengths for the magnetic vectors of the form
\begin{equation}
\label{MagneticBianchis_app}
\begin{array}{lll}
\tilde{H}^{\prime}_{\2 i } &=& d \tilde{A}^{\prime}_{i} + \tfrac12 \, g  \, \epsilon_{ij}{}^k  \, A^{\prime  j } \wedge \tilde{A}^{\prime}_ {k } + \tfrac12 \, g \,  \epsilon_{ij}{}^k  \, A^{ (\textrm{t})  j } \wedge \tilde{A}^{(\textrm{t}) }_ {k } - \tfrac12 \, m \,  \epsilon_{i}{}^{jk}  \,  \tilde{A}^{(\textrm{t}) }_ {j} \wedge \tilde{A}^{(\textrm{t}) }_ {k }  + g  \, \epsilon_{ij}{}^k   \,  B_k{}^j  \ , \\[2mm]
\tilde{H}^{(\textrm{L})  }_{\2  a}  &=& d  \tilde{A}^{(\textrm{L})  }_{ a}  + \tfrac12  \, g \, \epsilon_{a b }{}^{ c } \, A^{ (\textrm{L})  b  } \wedge \tilde{A}^{(\textrm{L})}_{c }  + g \,  B_{a} \ , \\[2mm]
\tilde{H}^{( \textrm{t})}_{ \2 i }  & = &  d \tilde{A}^{( \textrm{t})}_{ i }  + \tfrac12 \, g \, \epsilon_{ij}{}^k \, A^{\prime  j } \wedge \tilde{A}^{( \textrm{t})}_k  + g \, \delta_{ij} B^j \ ,
\end{array}
\end{equation} 
with $\,B_{i}{}^{j} = {B^{\scaleto{(A)}{6pt}}}_{i}{}^{j} + {B^{\scaleto{(S)}{6pt}}}_{i}{}^{j}\,$ and $\,B_{a}\,$ being the additional two-form potentials in (\ref{two-form_fields}). The magnetic field strengths in (\ref{MagneticBianchis_app}) are also subject to Bianchi identities involving covariant derivatives defined as
\begin{equation}
\label{DH_mag_app}
\begin{array}{lll}
D \tilde{H}^{\prime}_{\2 i } & \equiv & d \tilde{H}^{\prime}_{\2 i }   + g \, \epsilon_{ij}{}^k \, A^{\prime  j } \wedge \tilde{H}^{\prime}_{\2 k }  + \epsilon_{ij}{}^k \big( g A^{(\textrm{t}) j } - m \, \delta^{jh} \tilde{A}^{(\textrm{t}) }_h  \big) \wedge   \tilde{H}^{( \textrm{t})}_{\2 k } \ , \\[2mm]
D  \tilde{H}^{(\textrm{L})  }_{\2 a}   & \equiv &  d  \tilde{H}^{(\textrm{L})  }_{\2 a} + g \, \epsilon_{a b }{}^{ c  } \, A^{ (\textrm{L})  b} \wedge \tilde{H}^{(\textrm{L})  }_{\2  c} \ , \\[2mm]
D \tilde{H}^{( \textrm{t})}_{ \2 i }  & \equiv &  d  \tilde{H}^{( \textrm{t})}_{ \2 i }   +  g \, \epsilon_{ij}{}^k \, A^{\prime  j } \wedge  \tilde{H}^{( \textrm{t})}_{ \2 k }  \ ,
\end{array}
\end{equation} 
together with the electric charge $\,g\,$ and the three-form field strengths $\,H_{\3 i}{}^j\,$ and $\,H_{\3 a}\,$ of the two-form potentials $\,B_{i}{}^{j} \,$ and $\,B_{a}\,$ in (\ref{two-form_fields}). 

The various three-form field strengths above are connected with scalar currents via a set of duality relations. When restricted to the $\,\textrm{SO}(3)_{\textrm{R}}$-invariant sector, the tensor-scalar duality relations of \cite{Guarino:2015qaa} reduce to 
{\setlength\arraycolsep{2pt} 
\begin{eqnarray} \label{eq:ThreeFormFS1}
H_{\3 i}{}^j & =& \tfrac17 \, * \Big[ 4 \big( d\varphi - e^{2\varphi} \chi \, d\chi \big) -3 \sqrt{2}  \big( d \phi_1 + d \phi_2 + d \phi_3 \big) \nonumber \\[4pt]
&& \qquad -\tfrac12 \, e^{\sqrt{2} \, ( \phi_1 + \phi_2 + \phi_3 ) } \,  \big( 2 \,  \textrm{tr} \, \big( \bm{m} \bm{f} \bm{a}  \big)  - \textrm{tr} ( \bm{m} ) \, \textrm{tr} \, (  \bm{f} \bm{a} )  \big) 
+ 3\,  \textrm{tr} \, \big(  \bm{m}^{-1} \, \bm{b}^\textrm{T} \, D\bm{b} \big) \Big] \, \delta_i{}^j \nonumber \\[4pt]
&& - * ( \bm{m}^{-1} D \bm{m})_i{}^j  + \tfrac12 \, e^{\sqrt{2} \, ( \phi_1 + \phi_2 + \phi_3 ) }  *  \big( 2 \,  \big( \bm{f} \bm{a} \bm{m}  \big)_i{}^j   - \textrm{tr} \, (  \bm{f} \bm{a} ) \, m_i{}^j  \big)  \nonumber \\[4pt]
&& + \tfrac14 \, e^{\sqrt{2} \, ( \phi_1 + \phi_2 + \phi_3 ) }  \epsilon^{ipq} \epsilon^{kh\ell} \delta_{ab} \, b^a{}_j b^b{}_p \, m_{kq} * f_{h\ell} \nonumber \\[4pt]
&& - * \big(  \bm{b}^\textrm{T} \, D\bm{b} \, \bm{m}^{-1} \big)_i{}^j  +2 \, \delta_{ab} \, \delta^{jh} \, b^a{}_{[i} (m^{-1})_{h]}{}^k * D b^b{}_k \; ,  \\[12pt]
\label{eq:ThreeFormFS2}
H_\3^i  &=&  \tfrac12 \, e^{\sqrt{2} \, ( \phi_1 + \phi_2 + \phi_3 ) } \, m^i{}_j \,  \epsilon^{jkh} * f_{kh}  \; ,
\\[12pt]
\label{eq:ThreeFormFS3}
H_{\3 a} & =&  2 \, \epsilon_{abc} \, (m^{-1})^{ij}  \,  b^b{}_i  *  D b ^c{}_j - \tfrac12  \,  e^{\sqrt{2} \, ( \phi_1 + \phi_2 + \phi_3 ) } \, \epsilon_{abc} \, \epsilon^{ijk} \, \epsilon^{h \ell m} \, b^b{}_i \, b^c{}_j \, m_{kh} \, * f_{\ell m } \; .
\end{eqnarray}
}Equation (\ref{eq:ThreeFormFS2}) is just (\ref{3FormDualityRelations}) written out explicitly in the parameterisation that we are using. The dualisation conditions (\ref{eq:ThreeFormFS1})--(\ref{eq:ThreeFormFS3}) further reduce to the expressions given in \cite{Guarino:2015qaa} for the $\,\textrm{SU}(3)$-invariant sector upon the identifications in (\ref{N=4toSU3inv})-(\ref{N=4toSU3invVecsMagnetic}), as well as to the expressions given in \cite{DeLuca:2018buk} for the \mbox{$\,\textrm{SO}(4)_{\textrm{d} \times \textrm{R}}$-invariant} sector upon the identifications in (\ref{N=4toSO4inv}). Finally, the expression for the various three-form field strengths in terms of the corresponding two-form gauge potentials (and also vector fields) can be obtained by particularising the general expressions in eq.~$(2.8)$ of \cite{Guarino:2015qaa}.

The three-form potentials $\,C^{ij} \equiv (\textbf{5}+\textbf{1},\textbf{1})\,$ and $\,C^{0} \equiv (\textbf{1},\textbf{1})\,$ in (\ref{three-forms_electric}) dual to electric embedding tensor deformations have field strengths given by
{\setlength\arraycolsep{2pt}
\begin{eqnarray}
\label{H4_elec_1}
 \bm{H}_\4 &= & g \,  \textrm{vol}_4 \, \Big[ \Big( 4 \, e^{ \frac{1}{\sqrt{2} } (\phi_1 +\phi_2 +\phi_3 )} +\sqrt{2} \,  \chi \,  e^{\varphi +\sqrt{2} (\phi_1 +\phi_2 +\phi_3 )} \, \textrm{det} \, \bm{b} \Big) \big( \bm{m} +\tfrac12 \, \bm{b}^\textrm{T} \bm{b} \big)    \nonumber \\[2mm]
 && \quad + e^{-\varphi +\sqrt{2} (\phi_1 +\phi_2 +\phi_3 )}  \big(1 + e^{2\varphi} \chi^2 \big) \Big( ( \textrm{tr} \, \bm{m} ) \, \bm{m} -2 \bm{m} \bm{m} +\tfrac12 \, \bm{m} \, \bm{b}^\textrm{T} \bm{b}  +\tfrac12 \, \bm{b}^\textrm{T} \bm{b}  \, \bm{m}  
 \nonumber \\[2mm]
 &&\quad  +\tfrac14 \, \big( \textrm{tr} \,(  \bm{b}^\textrm{T} \bm{b} ) \big) \, \bm{b}^\textrm{T} \bm{b}  \Big) \Big] 
  +  \tfrac12 m \,  \chi \,  e^{\varphi +\sqrt{2} (\phi_1 +\phi_2 +\phi_3 )} \, \bm{b}^\textrm{T} \bm{b}  \, \textrm{vol}_4  \ , 
\end{eqnarray}
}and
{\setlength\arraycolsep{2pt}
\begin{eqnarray}
\label{H4_elec_2}
H^{0}_\4 & = & g \, \textrm{vol}_4 \, \Big[ \Big( 2 \, e^{ \frac{1}{\sqrt{2} } (\phi_1 +\phi_2 +\phi_3 )} + \tfrac{1}{2\sqrt{2}}  \,  \chi \,  e^{\varphi +\sqrt{2} (\phi_1 +\phi_2 +\phi_3 )} \, \textrm{det} \, \bm{b} \Big)  \,  \tr \, \big(  {\bm m} + \tfrac12 \, {\bm b}^\textrm{T} {\bm b}  \big) 
\nonumber \\[2mm]
&& \quad -\tfrac14 e^\varphi  \big( \tr \, ( {\bm b}^\textrm{T} {\bm b}  \,  {\bm m}^{-1} ) \big)^2   +  \tfrac14 e^\varphi  \,  \tr  ( {\bm b}^\textrm{T} {\bm b} \,  {\bm m}^{-1}  {\bm b}^\textrm{T} {\bm b}  \, {\bm m}^{-1}  ) - \tfrac12 \, e^{\varphi +\sqrt{2} ( \phi_1+  \phi_2+  \phi_3)} (\det  {\bm b} )^2 \Big] 
 \nonumber \\[2mm]
 && 
-  \tfrac{1}{2\sqrt{2}} \,  m  \,  e^{\varphi +\sqrt{2} (\phi_1 +\phi_2 +\phi_3 )} \, \textrm{det} \, \bm{b}  \, \textrm{vol}_4 \ .
\end{eqnarray}
}The expression for the various four-form field strengths in terms of the corresponding three-form gauge potentials (and also two-form and vector fields) can be obtained by particularising the general expressions in eq.~$(2.9)$ of \cite{Guarino:2015qaa}.

On the other hand, the three-form potential $\,\tilde{C} \equiv (\textbf{1},\textbf{1})\,$ in (\ref{three-forms_magnetic}) dual to the magnetic embedding tensor deformation has a field strength given by
\begin{eqnarray}
\label{H4_mag}
\tilde{H}_\4  = \Big[ \tfrac12 \, g  \, e^{\varphi +\sqrt{2} (\phi_1 +\phi_2 +\phi_3 )} \, \Big( \chi  \,  \tr \, \big( {\bm b}^\textrm{T} {\bm b}   \big) -2\sqrt{2} \, \textrm{det} \, \bm{b}   \Big) -  m  \,  e^{\varphi +\sqrt{2} (\phi_1 +\phi_2 +\phi_3 )} \Big] \, \textrm{vol}_4 \ .
\end{eqnarray}
Then the consistency of the $\,\textrm{SO}(3)_{\textrm{R}}$-invariant sector guarantees that the scalar potential in (\ref{V_SO3}) can be expressed in terms of the field strengths in (\ref{H4_elec_1})-(\ref{H4_elec_2}) and (\ref{H4_mag}). Indeed one finds that
\begin{equation}
\label{VfromH4}
g \, \big( \delta_{ij} \, H^{ij}_\4 + 4 \, H^{0}_\4 \big) + m  \, \tilde{H}_\4  = -2 \, V \, \textrm{vol}_4 \ ,
\end{equation}
in agreement with eqs.~($2.28$) and ($2.29$) of \cite{Guarino:2015qaa}.

Finally, as discussed in full generality for the $\,\textrm{ISO}(7)\,$ theory in \cite{Guarino:2015qaa}, substituting the duality relations into the Bianchi identities for the $\,\textrm{SO}(3)_{\textrm{R}}$-invariant hierarchy of fields, one obtains a projection of the scalar equations of motion. In terms of representations of $\,\textrm{SO}(3)_{\textrm{d}} \times \textrm{SO}(3)_{\textrm{R}} \subset \textrm{SO}(7)\,$ such scalar equations of motion are given by
\begin{equation} 
\label{NecCondVac_singlet}
(\bm{1},\bm{1}) \subset \bm{1} : \hspace{3mm}  g \, \big( \delta_{ij} \, H^{ij}_\4 + 4 \, H^{0}_\4 \big) + 7 m  \, \tilde{H}_\4  = 0 \ ,
\end{equation}
together with
\begin{equation} 
\label{NecCondVac_non-singlet}
\begin{array}{lrl}
(\bm{5} \,+\, \bm{1},\bm{1}) \subset \bm{27} : & \hspace{3mm} 7 \, H^{ij}_\4 - \big(  \delta_{hk} \, H^{hk}_\4 \, + \, 4 \, H^{0}_\4 \big) \, \delta^{ij} = 0 & , \\[2mm]
& \delta_{ij} \, H^{ij}_\4  \,- \, 3 \, H^{0}_\4 = 0 &  .
\end{array}
\end{equation}

\bibliography{references}

\providecommand{\href}[2]{#2}\begingroup\raggedright\begin{thebibliography}{10}

\bibitem{Maldacena:1997re}
J.~M. Maldacena, {\it {The Large N limit of superconformal field theories and
  supergravity}},  {\em Int. J. Theor. Phys.} {\bf 38} (1999) 1113--1133,
  [\href{http://arxiv.org/abs/hep-th/9711200}{{\tt hep-th/9711200}}]. [Adv.
  Theor. Math. Phys.2,231(1998)].

\bibitem{Pernici:1984xx}
M.~Pernici, K.~Pilch, and P.~van Nieuwenhuizen, {\it {Gauged Maximally Extended
  Supergravity in Seven-dimensions}},  {\em Phys. Lett.} {\bf B143} (1984) 103.

\bibitem{Gunaydin:1984qu}
M.~Gunaydin, L.~Romans, and N.~Warner, {\it {Gauged N=8 Supergravity in
  Five-Dimensions}},  {\em Phys.Lett.} {\bf B154} (1985) 268.

\bibitem{deWit:1982ig}
B.~de~Wit and H.~Nicolai, {\it {N=8 Supergravity}},  {\em Nucl.Phys.} {\bf
  B208} (1982) 323.

\bibitem{Aharony:2008ug}
O.~Aharony, O.~Bergman, D.~L. Jafferis, and J.~Maldacena, {\it {N=6
  superconformal Chern-Simons-matter theories, M2-branes and their gravity
  duals}},  {\em JHEP} {\bf 10} (2008) 091,
  [\href{http://arxiv.org/abs/0806.1218}{{\tt arXiv:0806.1218}}].

\bibitem{Guarino:2015jca}
A.~Guarino, D.~L. Jafferis, and O.~Varela, {\it {The string origin of dyonic
  N=8 supergravity and its simple Chern-Simons duals}},  {\em Phys. Rev. Lett.}
  {\bf 115} (2015), no.~9 091601, [\href{http://arxiv.org/abs/1504.08009}{{\tt
  arXiv:1504.08009}}].

\bibitem{Dall'Agata:2012bb}
G.~Dall'Agata, G.~Inverso, and M.~Trigiante, {\it {Evidence for a family of
  SO(8) gauged supergravity theories}},  {\em Phys.Rev.Lett.} {\bf 109} (2012)
  201301, [\href{http://arxiv.org/abs/1209.0760}{{\tt arXiv:1209.0760}}].

\bibitem{Dall'Agata:2014ita}
G.~Dall'Agata, G.~Inverso, and A.~Marrani, {\it {Symplectic Deformations of
  Gauged Maximal Supergravity}},  {\em JHEP} {\bf 1407} (2014) 133,
  [\href{http://arxiv.org/abs/1405.2437}{{\tt arXiv:1405.2437}}].

\bibitem{Inverso:2015viq}
G.~Inverso, {\it {Electric-magnetic deformations of D = 4 gauged
  supergravities}},  {\em JHEP} {\bf 03} (2016) 138,
  [\href{http://arxiv.org/abs/1512.04500}{{\tt arXiv:1512.04500}}].

\bibitem{Hull:1984yy}
C.~Hull, {\it {New Gauging of $N=8$ Supergravity}},  {\em Phys.Rev.} {\bf D30}
  (1984) 760.

\bibitem{Hull:1988jw}
C.~Hull and N.~Warner, {\it {Noncompact Gaugings From Higher Dimensions}},
  {\em Class.Quant.Grav.} {\bf 5} (1988) 1517.

\bibitem{Giani:1984wc}
F.~Giani and M.~Pernici, {\it {N=2 SUPERGRAVITY IN TEN-DIMENSIONS}},  {\em
  Phys. Rev.} {\bf D30} (1984) 325--333.

\bibitem{DallAgata:2011aa}
G.~Dall'Agata and G.~Inverso, {\it {On the Vacua of N = 8 Gauged Supergravity
  in 4 Dimensions}},  {\em Nucl.Phys.} {\bf B859} (2012) 70--95,
  [\href{http://arxiv.org/abs/1112.3345}{{\tt arXiv:1112.3345}}].

\bibitem{Guarino:2015qaa}
A.~Guarino and O.~Varela, {\it {Dyonic ISO(7) supergravity and the duality
  hierarchy}},  {\em JHEP} {\bf 02} (2016) 079,
  [\href{http://arxiv.org/abs/1508.04432}{{\tt arXiv:1508.04432}}].

\bibitem{Guarino:2015vca}
A.~Guarino and O.~Varela, {\it {Consistent $ \mathcal{N}=8 $ truncation of
  massive IIA on S$^{6}$}},  {\em JHEP} {\bf 12} (2015) 020,
  [\href{http://arxiv.org/abs/1509.02526}{{\tt arXiv:1509.02526}}].

\bibitem{Ciceri:2016dmd}
F.~Ciceri, A.~Guarino, and G.~Inverso, {\it {The exceptional story of massive
  IIA supergravity}},  {\em JHEP} {\bf 08} (2016) 154,
  [\href{http://arxiv.org/abs/1604.08602}{{\tt arXiv:1604.08602}}].

\bibitem{Cassani:2016ncu}
D.~Cassani, O.~de~Felice, M.~Petrini, C.~Strickland-Constable, and D.~Waldram,
  {\it {Exceptional generalised geometry for massive IIA and consistent
  reductions}},  {\em JHEP} {\bf 08} (2016) 074,
  [\href{http://arxiv.org/abs/1605.00563}{{\tt arXiv:1605.00563}}].

\bibitem{Inverso:2016eet}
G.~Inverso, H.~Samtleben, and M.~Trigiante, {\it {Type II supergravity origin
  of dyonic gaugings}},  {\em Phys. Rev.} {\bf D95} (2017), no.~6 066020,
  [\href{http://arxiv.org/abs/1612.05123}{{\tt arXiv:1612.05123}}].

\bibitem{Romans:1985tz}
L.~Romans, {\it {Massive N=2a Supergravity in Ten-Dimensions}},  {\em
  Phys.Lett.} {\bf B169} (1986) 374.

\bibitem{Borghese:2012qm}
A.~Borghese, A.~Guarino, and D.~Roest, {\it {All $G_2$ invariant critical
  points of maximal supergravity}},  {\em JHEP} {\bf 1212} (2012) 108,
  [\href{http://arxiv.org/abs/1209.3003}{{\tt arXiv:1209.3003}}].

\bibitem{Gallerati:2014xra}
A.~Gallerati, H.~Samtleben, and M.~Trigiante, {\it {The $ \mathcal{N}>2 $
  supersymmetric AdS vacua in maximal supergravity}},  {\em JHEP} {\bf 12}
  (2014) 174, [\href{http://arxiv.org/abs/1410.0711}{{\tt arXiv:1410.0711}}].

\bibitem{Varela:2015uca}
O.~Varela, {\it {AdS$_{4}$ solutions of massive IIA from dyonic ISO(7)
  supergravity}},  {\em JHEP} {\bf 03} (2016) 071,
  [\href{http://arxiv.org/abs/1509.07117}{{\tt arXiv:1509.07117}}].

\bibitem{Pang:2015vna}
Y.~Pang and J.~Rong, {\it {N=3 solution in dyonic ISO(7) gauged maximal
  supergravity and its uplift to massive type IIA supergravity}},  {\em Phys.
  Rev.} {\bf D92} (2015), no.~8 085037,
  [\href{http://arxiv.org/abs/1508.05376}{{\tt arXiv:1508.05376}}].

\bibitem{DeLuca:2018buk}
G.~B. De~Luca, G.~L. Monaco, N.~T. Macpherson, A.~Tomasiello, and O.~Varela,
  {\it {The geometry of $ \mathcal{N}=3 $ AdS$_{4}$ in massive IIA}},  {\em
  JHEP} {\bf 08} (2018) 133, [\href{http://arxiv.org/abs/1805.04823}{{\tt
  arXiv:1805.04823}}].

\bibitem{Schwarz:2004yj}
J.~H. Schwarz, {\it {Superconformal Chern-Simons theories}},  {\em JHEP} {\bf
  11} (2004) 078, [\href{http://arxiv.org/abs/hep-th/0411077}{{\tt
  hep-th/0411077}}].

\bibitem{Gaiotto:2007qi}
D.~Gaiotto and X.~Yin, {\it {Notes on superconformal Chern-Simons-Matter
  theories}},  {\em JHEP} {\bf 08} (2007) 056,
  [\href{http://arxiv.org/abs/0704.3740}{{\tt arXiv:0704.3740}}].

\bibitem{Guarino:2016ynd}
A.~Guarino, J.~Tarrio, and O.~Varela, {\it {Romans-mass-driven flows on the
  D2-brane}},  {\em JHEP} {\bf 08} (2016) 168,
  [\href{http://arxiv.org/abs/1605.09254}{{\tt arXiv:1605.09254}}].

\bibitem{Pang:2015rwd}
Y.~Pang and J.~Rong, {\it {Evidence for the Holographic dual of ${\cal N}=3$
  Solution in Massive Type IIA}},  {\em Phys. Rev.} {\bf D93} (2016), no.~6
  065038, [\href{http://arxiv.org/abs/1511.08223}{{\tt arXiv:1511.08223}}].

\bibitem{Pang:2017omp}
Y.~Pang, J.~Rong, and O.~Varela, {\it {Spectrum universality properties of
  holographic Chern-Simons theories}},  {\em JHEP} {\bf 01} (2018) 061,
  [\href{http://arxiv.org/abs/1711.07781}{{\tt arXiv:1711.07781}}].

\bibitem{Bobev:2018ugk}
N.~Bobev, P.~Bomans, and F.~F. Gautason, {\it {Spherical Branes}},  {\em JHEP}
  {\bf 08} (2018) 029, [\href{http://arxiv.org/abs/1805.05338}{{\tt
  arXiv:1805.05338}}].

\bibitem{Suh:2018nmp}
M.~Suh, {\it {Supersymmetric Janus solutions of dyonic $ISO(7)$-gauged
  $\mathcal{N}\,=\,8$ supergravity}},  {\em JHEP} {\bf 04} (2018) 109,
  [\href{http://arxiv.org/abs/1803.00041}{{\tt arXiv:1803.00041}}].

\bibitem{Guarino:2017eag}
A.~Guarino and J.~Tarrío, {\it {BPS black holes from massive IIA on S$^{6}$}},
  {\em JHEP} {\bf 09} (2017) 141, [\href{http://arxiv.org/abs/1703.10833}{{\tt
  arXiv:1703.10833}}].

\bibitem{Hosseini:2017fjo}
S.~M. Hosseini, K.~Hristov, and A.~Passias, {\it {Holographic microstate
  counting for AdS$_{4}$ black holes in massive IIA supergravity}},  {\em JHEP}
  {\bf 10} (2017) 190, [\href{http://arxiv.org/abs/1707.06884}{{\tt
  arXiv:1707.06884}}].

\bibitem{Guarino:2017pkw}
A.~Guarino, {\it {BPS black hole horizons from massive IIA}},  {\em JHEP} {\bf
  08} (2017) 100, [\href{http://arxiv.org/abs/1706.01823}{{\tt
  arXiv:1706.01823}}].

\bibitem{Azzurli:2017kxo}
F.~Azzurli, N.~Bobev, P.~M. Crichigno, V.~S. Min, and A.~Zaffaroni, {\it {A
  universal counting of black hole microstates in AdS$_{4}$}},  {\em JHEP} {\bf
  02} (2018) 054, [\href{http://arxiv.org/abs/1707.04257}{{\tt
  arXiv:1707.04257}}].

\bibitem{Benini:2017oxt}
F.~Benini, H.~Khachatryan, and P.~Milan, {\it {Black hole entropy in massive
  Type IIA}},  {\em Class. Quant. Grav.} {\bf 35} (2018), no.~3 035004,
  [\href{http://arxiv.org/abs/1707.06886}{{\tt arXiv:1707.06886}}].

\bibitem{Liu:2018bac}
J.~T. Liu, L.~A. Pando~Zayas, and S.~Zhou, {\it {Subleading Microstate Counting
  in the Dual to Massive Type IIA}},
  \href{http://arxiv.org/abs/1808.10445}{{\tt arXiv:1808.10445}}.

\bibitem{Fluder:2015eoa}
M.~Fluder and J.~Sparks, {\it {D2-brane Chern-Simons theories: F-maximization =
  a-maximization}},  {\em JHEP} {\bf 01} (2016) 048,
  [\href{http://arxiv.org/abs/1507.05817}{{\tt arXiv:1507.05817}}].

\bibitem{Araujo:2016jlx}
T.~R. Araujo and H.~Nastase, {\it {Observables in the
  Guarino-Jafferis-Varela/CS-SYM duality}},  {\em JHEP} {\bf 07} (2017) 020,
  [\href{http://arxiv.org/abs/1609.08008}{{\tt arXiv:1609.08008}}].

\bibitem{Araujo:2017hvi}
T.~Araujo, G.~Itsios, H.~Nastase, and E.~Ó. Colgáin, {\it {Penrose limits and
  spin chains in the GJV/CS-SYM duality}},  {\em JHEP} {\bf 12} (2017) 137,
  [\href{http://arxiv.org/abs/1706.02711}{{\tt arXiv:1706.02711}}].

\bibitem{deWit:2013ija}
B.~de~Wit and H.~Nicolai, {\it {Deformations of gauged SO(8) supergravity and
  supergravity in eleven dimensions}},  {\em JHEP} {\bf 05} (2013) 077,
  [\href{http://arxiv.org/abs/1302.6219}{{\tt arXiv:1302.6219}}].

\bibitem{Lee:2015xga}
K.~Lee, C.~Strickland-Constable, and D.~Waldram, {\it {New Gaugings and
  Non-Geometry}},  {\em Fortsch. Phys.} {\bf 65} (2017), no.~10-11 1700049,
  [\href{http://arxiv.org/abs/1506.03457}{{\tt arXiv:1506.03457}}].

\bibitem{Cremmer:1978km}
E.~Cremmer, B.~Julia, and J.~Scherk, {\it {Supergravity Theory in
  Eleven-Dimensions}},  {\em Phys.Lett.} {\bf B76} (1978) 409--412.

\bibitem{deWit:1986iy}
B.~de~Wit and H.~Nicolai, {\it {The Consistency of the $S^7$ Truncation in
  $D=11$ Supergravity}},  {\em Nucl.Phys.} {\bf B281} (1987) 211.

\bibitem{Minwalla:2011ma}
S.~Minwalla, P.~Narayan, T.~Sharma, V.~Umesh, and X.~Yin, {\it {Supersymmetric
  States in Large N Chern-Simons-Matter Theories}},  {\em JHEP} {\bf 02} (2012)
  022, [\href{http://arxiv.org/abs/1104.0680}{{\tt arXiv:1104.0680}}].

\bibitem{deWit:2008ta}
B.~de~Wit, H.~Nicolai, and H.~Samtleben, {\it {Gauged Supergravities, Tensor
  Hierarchies, and M-Theory}},  {\em JHEP} {\bf 02} (2008) 044,
  [\href{http://arxiv.org/abs/0801.1294}{{\tt arXiv:0801.1294}}].

\bibitem{deWit:2008gc}
B.~de~Wit and H.~Samtleben, {\it {The End of the p-form hierarchy}},  {\em
  JHEP} {\bf 08} (2008) 015, [\href{http://arxiv.org/abs/0805.4767}{{\tt
  arXiv:0805.4767}}].

\bibitem{Bergshoeff:2009ph}
E.~A. Bergshoeff, J.~Hartong, O.~Hohm, M.~Huebscher, and T.~Ortin, {\it {Gauge
  Theories, Duality Relations and the Tensor Hierarchy}},  {\em JHEP} {\bf
  0904} (2009) 123, [\href{http://arxiv.org/abs/0901.2054}{{\tt
  arXiv:0901.2054}}].

\bibitem{Kim:2018sdw}
H.~Kim, N.~Kim, and M.~Suh, {\it {On the U(1)$^{2}$-Invariant Sector of Dyonic
  Maximal Supergravity}},  {\em J. Korean Phys. Soc.} {\bf 73} (2018), no.~3
  249--258, [\href{http://arxiv.org/abs/1801.01286}{{\tt arXiv:1801.01286}}].

\bibitem{Dibitetto:2011eu}
G.~Dibitetto, A.~Guarino, and D.~Roest, {\it {How to halve maximal
  supergravity}},  {\em JHEP} {\bf 06} (2011) 030,
  [\href{http://arxiv.org/abs/1104.3587}{{\tt arXiv:1104.3587}}].

\bibitem{Schon:2006kz}
J.~Schon and M.~Weidner, {\it {Gauged N=4 supergravities}},  {\em JHEP} {\bf
  05} (2006) 034, [\href{http://arxiv.org/abs/hep-th/0602024}{{\tt
  hep-th/0602024}}].

\bibitem{deWit:2005ub}
B.~de~Wit, H.~Samtleben, and M.~Trigiante, {\it {Magnetic charges in local
  field theory}},  {\em JHEP} {\bf 0509} (2005) 016,
  [\href{http://arxiv.org/abs/hep-th/0507289}{{\tt hep-th/0507289}}].

\bibitem{deRoo:1985jh}
M.~de~Roo and P.~Wagemans, {\it {Gauge Matter Coupling in $N=4$ Supergravity}},
   {\em Nucl. Phys.} {\bf B262} (1985) 644. [,666(1985)].

\bibitem{deRoo:2002jf}
M.~de~Roo, D.~B. Westra, and S.~Panda, {\it {De Sitter solutions in N=4 matter
  coupled supergravity}},  {\em JHEP} {\bf 02} (2003) 003,
  [\href{http://arxiv.org/abs/hep-th/0212216}{{\tt hep-th/0212216}}].

\bibitem{deRoo:2003rm}
M.~de~Roo, D.~B. Westra, S.~Panda, and M.~Trigiante, {\it {Potential and mass
  matrix in gauged N=4 supergravity}},  {\em JHEP} {\bf 11} (2003) 022,
  [\href{http://arxiv.org/abs/hep-th/0310187}{{\tt hep-th/0310187}}].

\bibitem{deRoo:2006ms}
M.~de~Roo, D.~B. Westra, and S.~Panda, {\it {Gauging CSO groups in N=4
  Supergravity}},  {\em JHEP} {\bf 09} (2006) 011,
  [\href{http://arxiv.org/abs/hep-th/0606282}{{\tt hep-th/0606282}}].

\bibitem{Roest:2009tt}
D.~Roest and J.~Rosseel, {\it {De Sitter in Extended Supergravity}},  {\em
  Phys. Lett.} {\bf B685} (2010) 201--207,
  [\href{http://arxiv.org/abs/0912.4440}{{\tt arXiv:0912.4440}}].

\bibitem{Dibitetto:2011gm}
G.~Dibitetto, A.~Guarino, and D.~Roest, {\it {Charting the landscape of N=4
  flux compactifications}},  {\em JHEP} {\bf 1103} (2011) 137,
  [\href{http://arxiv.org/abs/1102.0239}{{\tt arXiv:1102.0239}}].

\bibitem{Dibitetto:2012ia}
G.~Dibitetto, A.~Guarino, and D.~Roest, {\it {Exceptional Flux
  Compactifications}},  {\em JHEP} {\bf 1205} (2012) 056,
  [\href{http://arxiv.org/abs/1202.0770}{{\tt arXiv:1202.0770}}].

\bibitem{Guarino:2015tja}
A.~Guarino, {\it {CSO$_c$ superpotentials}},  {\em Nucl. Phys.} {\bf B900}
  (2015) 501--516, [\href{http://arxiv.org/abs/1508.05055}{{\tt
  arXiv:1508.05055}}].

\bibitem{Fre:1999gok}
P.~Fre', L.~Gualtieri, and P.~Termonia, {\it {The Structure of N=3 multiplets
  in AdS(4) and the complete Osp(3|4) x SU(3) spectrum of M theory on AdS(4) x
  N0,1,0}},  {\em Phys. Lett.} {\bf B471} (1999) 27--38,
  [\href{http://arxiv.org/abs/hep-th/9909188}{{\tt hep-th/9909188}}].

\bibitem{Guarino:2019snw}
A.~Guarino, J.~Tarrio, and O.~Varela, {\it {Flowing to $\mathcal{N}=3$
  Chern--Simons-matter theory}},  \href{http://arxiv.org/abs/1910.06866}{{\tt
  arXiv:1910.06866}}.

\bibitem{Lu:1998xt}
H.~Lu, C.~N. Pope, and K.~S. Stelle, {\it {M theory / heterotic duality: A
  Kaluza-Klein perspective}},  {\em Nucl. Phys.} {\bf B548} (1999) 87--138,
  [\href{http://arxiv.org/abs/hep-th/9810159}{{\tt hep-th/9810159}}].

\bibitem{deWit:2007mt}
B.~de~Wit, H.~Samtleben, and M.~Trigiante, {\it {The Maximal D=4
  supergravities}},  {\em JHEP} {\bf 0706} (2007) 049,
  [\href{http://arxiv.org/abs/0705.2101}{{\tt arXiv:0705.2101}}].

\end{thebibliography}\endgroup

\end{document}